\documentclass[aps,prd,floatfix,showpacs]{revtex4}
\usepackage{amsmath,amsfonts}
\usepackage[dvips]{graphicx}

\begin{document}

\title{Constraints on oscillating dark energy models}
\author{Aleksandra Kurek}
\email{alex@oa.uj.edu.pl}
\affiliation{Astronomical Observatory, Jagiellonian University,
Orla 171, 30-244 Krak{\'o}w, Poland}
\author{Orest Hrycyna}
\email{hrycyna@kul.lublin.pl}
\affiliation{Department of Theoretical Physics, Faculty of Philosophy, 
The John Paul II Catholic University of Lublin, Al. Rac{\l}awickie 14, 20-950
Lublin, Poland}
\author{Marek Szyd{\l}owski}
\email{uoszydlo@cyf-kr.edu.pl}
\affiliation{Astronomical Observatory, Jagiellonian University,
Orla 171, 30-244 Krak{\'o}w, Poland}
\affiliation{Mark Kac Complex Systems Research Centre, Jagiellonian University,
Reymonta 4, 30-059 Krak{\'o}w, Poland}


\begin{abstract}
The oscillating scenario of route to Lambda was recently proposed by us
\cite{Hrycyna:2007mq} as an alternative to a cosmological constant in a
explanation of the current accelerating universe. In this scenario phantom
scalar field conformally coupled to gravity drives the accelerating phase of the
universe. In our model $\Lambda$CDM appears as a global attractor in the phase
space. In this paper we investigate observational constraints on this scenario
from recent measurements of distant supernovae type Ia, $H(z)$ observational
data, CMB R shift and BAO parameter.
The Bayesian methods of model selection are used in comparison the model with
concordance $\Lambda$CDM one as well as with model with dynamical dark energy
parametrised by linear form. We conclude that $\Lambda$CDM is 
favoured over FRW model with dynamical oscillating dark energy. Our analysis
also demonstrate that FRW model with oscillating dark energy is favoured over
FRW model with decaying dark energy parametrised in linear way.
\end{abstract}

\pacs{98.80.Es, 98.80.Cq, 95.36.+x}

\maketitle

\section{Introduction}

Observations of distant supernovae type Ia still consistently suggest that the
universe is in a accelerating phase of expansion
\cite{Astier:2005qq,Riess:1998cb,Davis:2007na}. These confirmations are
supported by CMB observations which indicate that universe is almost spatially
flat \cite{Spergel:2006hy} and that the amount of matter in the universe
calculated from galaxy clustering is not enough to account for this flatness
\cite{Cole:2005sx,Tegmark:2003ud}. These observational facts regarded on the
background of standard general relativity indicate that about $2/3$ of total
energy of the universe today being a dark energy with negative pressure which is
responsible for the current accelerated expansion if the strong energy condition
is violated.

There are many candidates for dark energy description \cite[and references
therein]{Copeland:2006wr}. Here we consider dark energy in the form of phantom
scalar field $\psi$ with the quadratic potential function $U(\psi)$ for simplicity
of presentation. The scalar field is conformally coupled to gravity. In our
previous work it has been demonstrated that for generic class of initial
conditions the equation of state parameter
$w_{\text{eff}}=p_{\text{eff}}/\rho_{\text{eff}}$ approaches $-1$ value through
the damping oscillations around this mysterious value. Hence theoretically
appeared the possibility to solve the cosmological constant problem where the
smallness of of cosmological constant does not require fine tuning of model
parameters.

Here we use different astronomical observations to confront the model
with the observational data. In this paper we use SNIa data and other tests like
CMB R shift, BAO and $H(z)$ data obtained from differential ages of galaxies
\cite{Simon:2004tf}. Bayesian statistics is used to constrain a set of model
parameters. In the constraining the model parameters we perform combined
analysis with CMB R shift parameter as calculated by Wang and Mukherjee
\cite{Wang:2006ts} for WMAP 3 \cite{Spergel:2006hy}. The main question addressing
in this paper is whether data sets to favour an evolving in oscillatory way dark
energy model over $\Lambda$CDM one. Using Bayesian framework of model selection
we also compare oscillating parametrisation with other most popular linear in
scale factor $a$ parametrisation.

Guo, Ohta and Zhang \cite{Guo:2005at} developed theoretical method of
reconstruction of the
quintessence potential directly from the effective equation of state parameter
$w(z)$ for minimally coupled scalar field. This method can be extended to the case of
non-minimally coupled scalar field.

\section{Oscillating dark energy model}

Investigations of different dark energy models \cite{Copeland:2006wr} are
hindered by lack of alternatives to the effective cosmological constant model
\cite{Crittenden:2007yy}. The simple step toward more realistic description is
that the dark energy might vary in time. Usually the form of $w(z)$ is a priori
assumption to remove some degeneration problem in analysis of constraints on
model parameters from observational data. However may happened that assumed form
of parametrisation of the dark energy equation of state is incompatible with
true dynamics which determine $w(z)$ itself. We propose to determine
corresponding form of $w(z)$ directly from the dynamical behaviour in the
vicinity of stable critical point representing effective model $\Lambda$CDM.
From the dynamical systems methods we know that the system in the phase space
can be good approximated by its linear part \cite{Perko:1991}. Then we solve
differential equation determining $w_{X}(z)$. As a result we obtain
\cite{Hrycyna:2007mq}
\begin{equation}
w_{X}(z)= -1 + (1+z)^{3}\bigg\{ C_{1}\cos(\ln(1+z)) + C_{2}\sin(\ln(1+z))\bigg\}
\label{eq:1}
\end{equation}
for phantom scalar field non-minimally (conformally) coupled to gravity
\cite{Faraoni:2000gx}. Note that a single scalar field model with general
Lagrangian $L=L(\phi,\partial_{\mu}\phi\partial^{\mu}\phi)$ will not be able to
have $w$ crossing $-1$ \cite{Bonvin:2006vc} and to realize that one must introduce
non-minimal coupling or modification of Einstein gravity.

We consider conformally coupled phantom scalar field with $p_{\psi}$ and
$\rho_{\psi}$ given by

\begin{align}
p_{\psi} & =-\frac{1}{2}\dot{\psi}^{2} - U(\psi) +\xi \Big[ 2H(\psi^{2})\dot{} +
(\psi^{2})\ddot{} \Big] + \xi \Big[ 2\dot{H}+3H^{2} \Big]\psi^{2},  \nonumber \\
\rho_{\psi} & = -\frac{1}{2}\dot{\psi}^{2} + U(\psi) - 3\xi H^{2} \psi^{2} - 3\xi
H(\psi^{2})\dot{}, \nonumber
\end{align}
where dot denotes differentiation with respect to cosmological time.

From eq.(\ref{eq:1}), instead of most popular linear parametrisation, we obtain
model with characteristic crossing of $w_{X}=-1$ ``phantom divide'', thereby the
violation of weak energy condition infinite times in the past.

With the help of formula (\ref{eq:1}) one can simply calculate energy density
for dark energy $\rho_{\Lambda}$
\begin{equation}
\rho_{\Lambda}=\rho_{\Lambda,0}\exp(-D_2)
\exp{\bigg((1+z)^{3}\Big[D_1\sin(\ln(1+z))+D_2\cos(\ln(1+z))\Big]\bigg)},
\label{eq:2}
\end{equation}
where $D_1=0.3(C_1+3C_2)$ and $D_2=0.3(3C_1-C_2)$.
It is interesting that some special cases of this dark energy parametrisation
are explored in probing for dynamics of dark energy \cite{Zhao:2006qg,
Hooper:2005xx}.

Let us consider flat FRW model filled with dark energy with density
$\rho_{\Lambda}$, dust matter (baryonic and dark) and radiation. For further
analysis of constraints from cosmography it would be useful to write Friedmann
first integral on $H^{2}$, where $H$ is the Hubble's parameter
\begin{equation}
H =
H_{0}\sqrt{\Omega_{\Lambda,0}\exp(-D_2)\exp{\bigg((1+z)^{3}\Big[D_1\sin(\ln(1+z))+D_2\cos(\ln(1+z))\Big]\bigg)}
+\Omega_{m,0}(1+z)^{3} + \Omega_{r,0}(1+z)^{4}},
\label{eq:3}
\end{equation}
where $\Omega_{r,0}\simeq 0.5 * 10^{-4}$ is fixed and
$\Omega_{\Lambda,0}=1-\Omega_{m,0}-\Omega_{r,0}$. $D_1$, $D_2$ and $\Omega_{m,0}$ are free
parameters which should be fitted from observational data.

\section{Constraints from SNIa, SDSS, CMB and H(z) observations}

To constrain the unknown values of model parameters we used the set of $N=192$
SNIa data \cite{Riess:2006fw, Davis:2007na, WoodVasey:2007jb}. Here we based on
the standard relation between the apparent magnitude ($m$) and luminosity
distance ($d_L$): $m-M=5\log _{10} D_L + \mathcal{M}$, where $M$ is the absolute
magnitude of SNIa, $\mathcal{M}=-5\log _{10} H_0 +25$ and $D_L=H_0 d_L$. The
luminosity distance depends on the considered cosmological model and with
assumption that $k=0$ is given by $d_{L}=(1+z)c\int_{0}^{z} \frac{d z'}{H(z')}$. \\
Posterior probability for model parameters (after marginalization over nuisance
parameter - $H_0$ with the assumption that prior probability for this parameter is flat within the interval $<60,80>$) has the following form
\begin{equation}
P(\bar{\theta}|M,D) \propto \int \big ( \pi(\bar{\theta}|M) \exp \left[-0.5\chi ^2_{SN}(\bar{\theta}) \right] \big ) dH_0,
\end{equation}
where $\chi ^2_{SN}(\bar{\theta})=\sum _{i=1}^{N} \left (\frac{\mu_i^{obs}-\mu_i^{th}}{\sigma_i} \right )^2$, $\mu_{i}^{obs}=m_{i}-M$, $\mu_{i}^{th}=5\log_{10}D_{Li} + \mathcal{M}$, $\pi(\bar{\theta}|M)$ is the prior probability for model parameters and $\bar{\theta} = (\Omega_{m,0},D_1,D_2)$. Here we assumed flat prior for model parameters within the interval: $\Omega_{m,0} \in < 0,1 >$, $D_1 \in <-1,1 >$, $D_2 \in <-1,1>$. \\

\begin{figure}[h]
\includegraphics[scale=0.5]{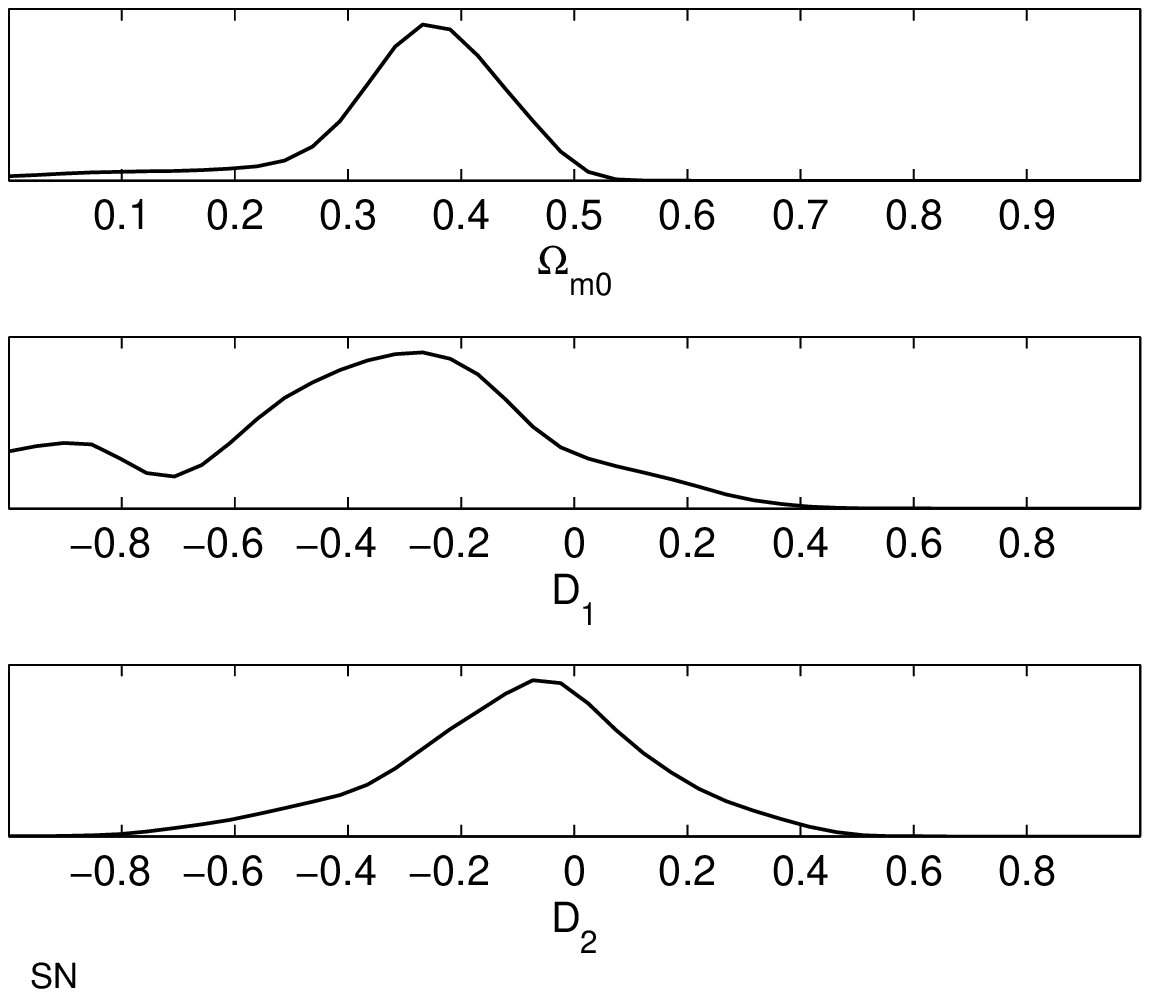}
\includegraphics[scale=0.5]{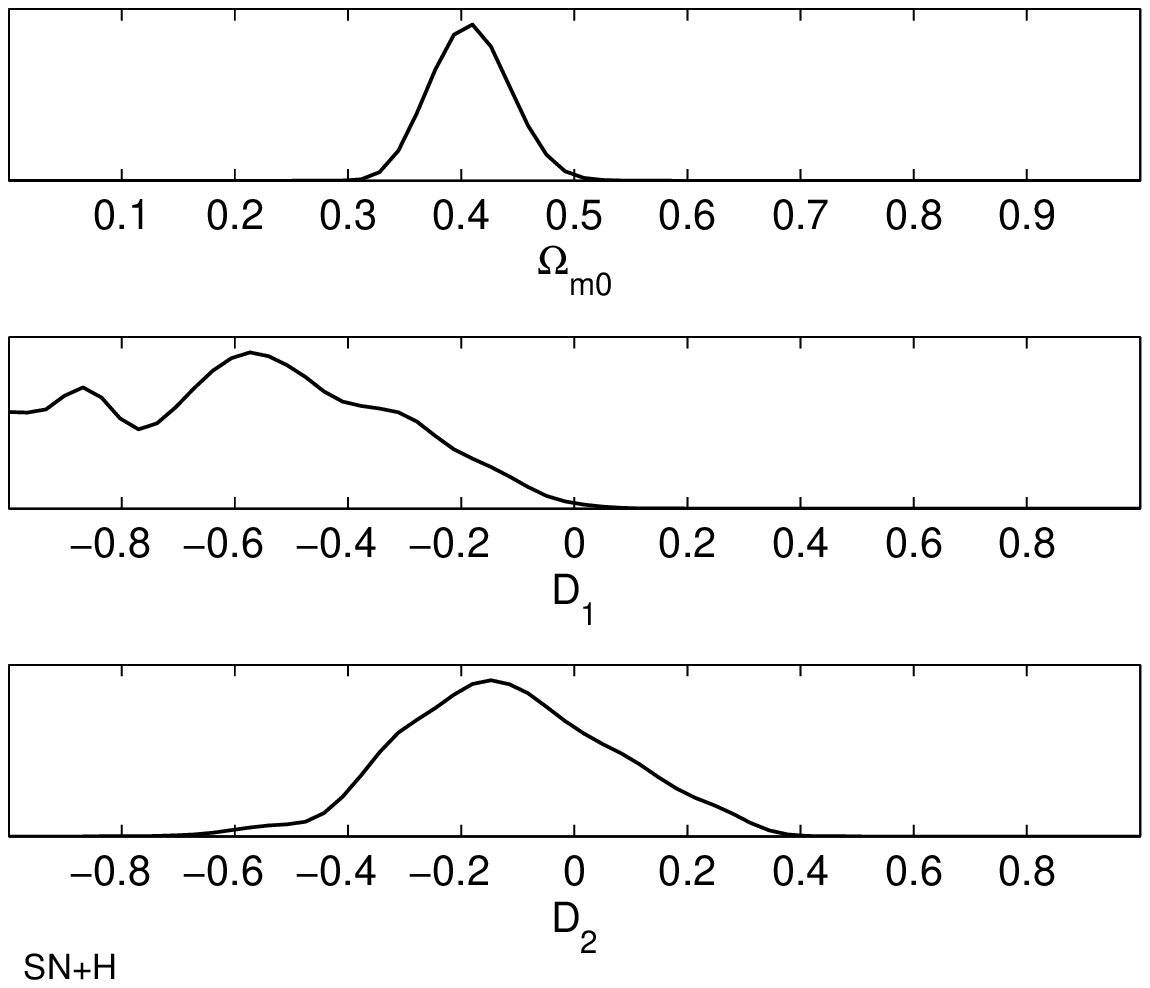}
\includegraphics[scale=0.5]{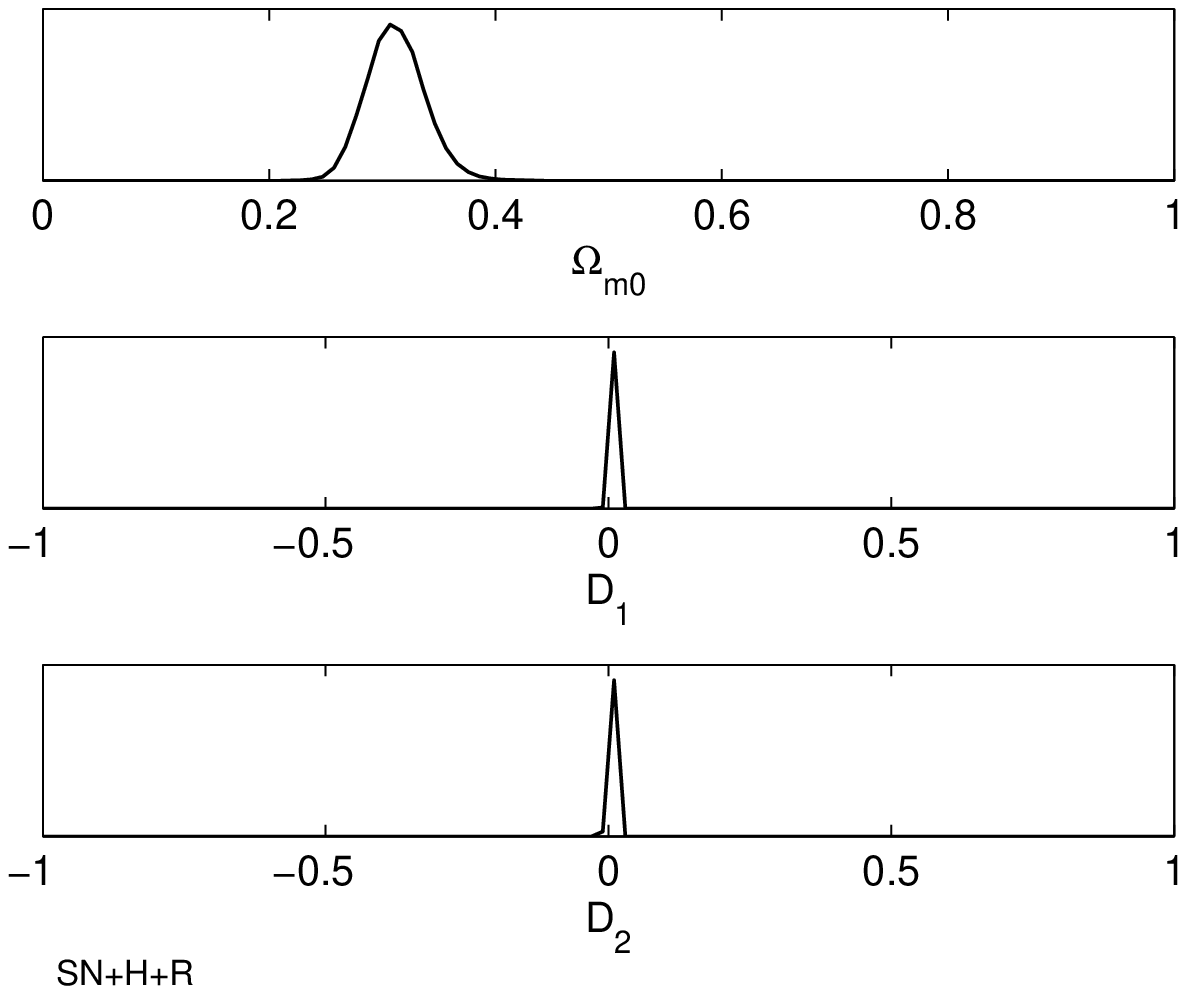}
\includegraphics[scale=0.5]{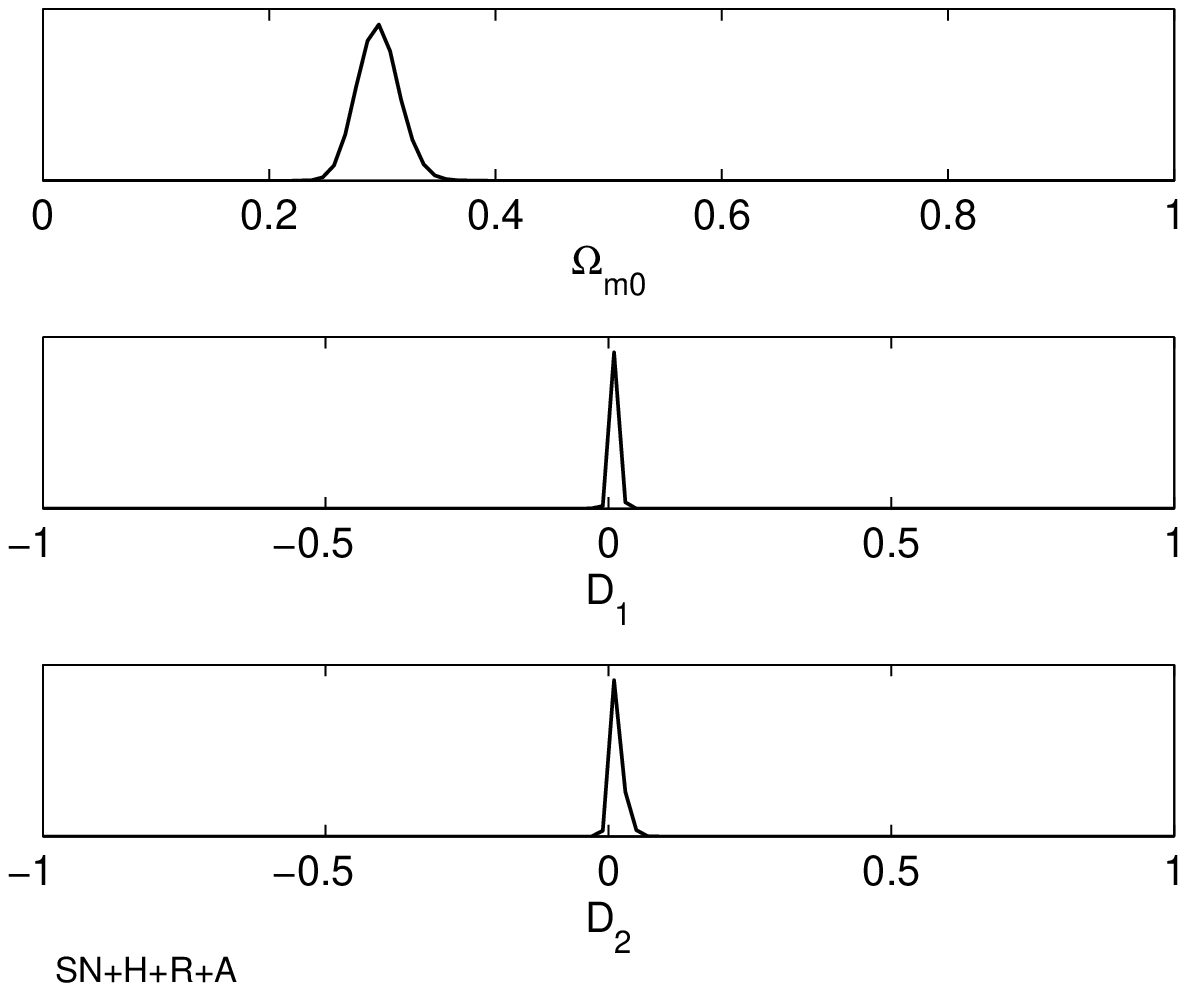}
\caption{Posterior probability distributions for oscillating DE model parameters.}
\label{fig:1}
\end{figure}

\begin{figure}[h]
\includegraphics[scale=0.5]{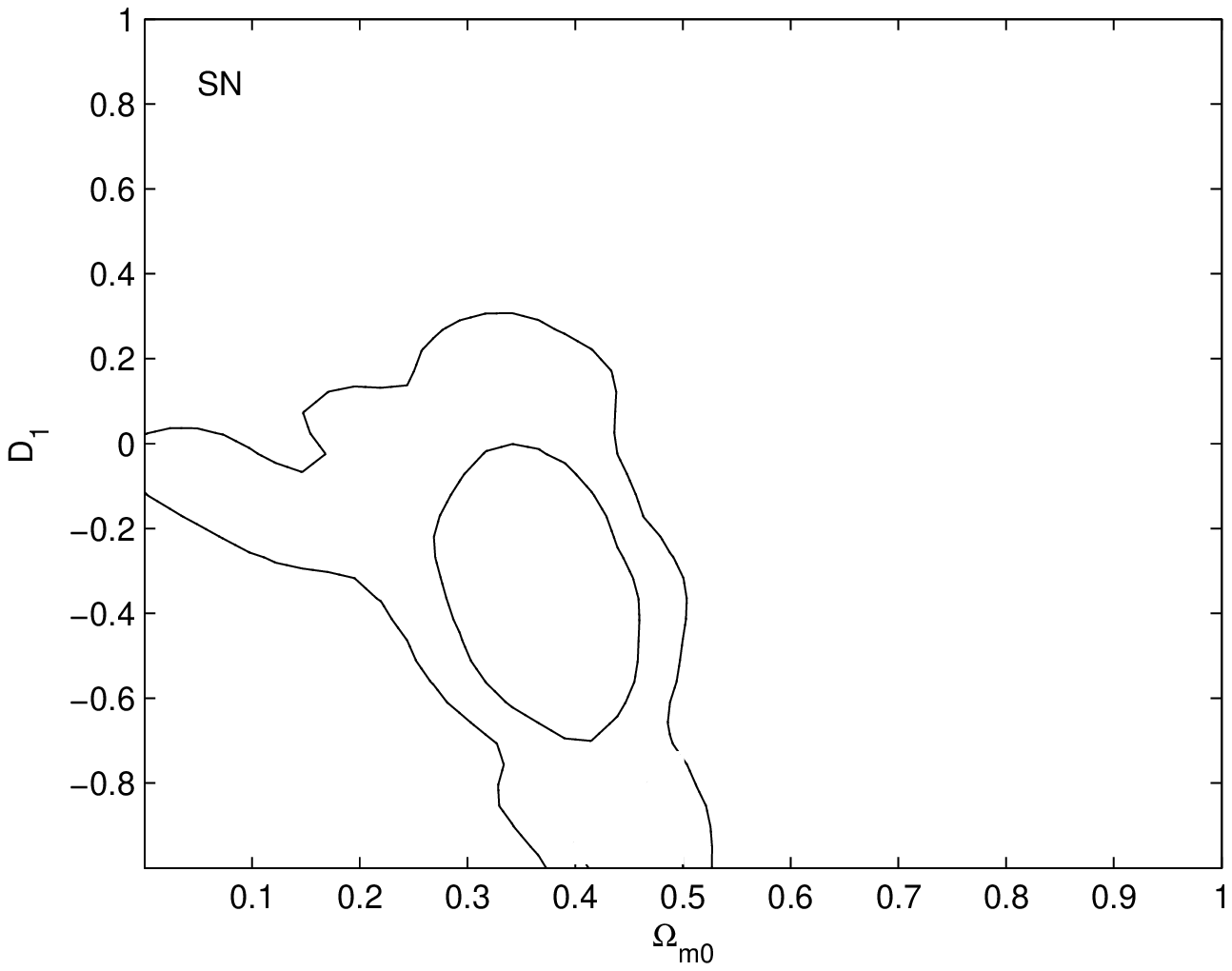}
\includegraphics[scale=0.5]{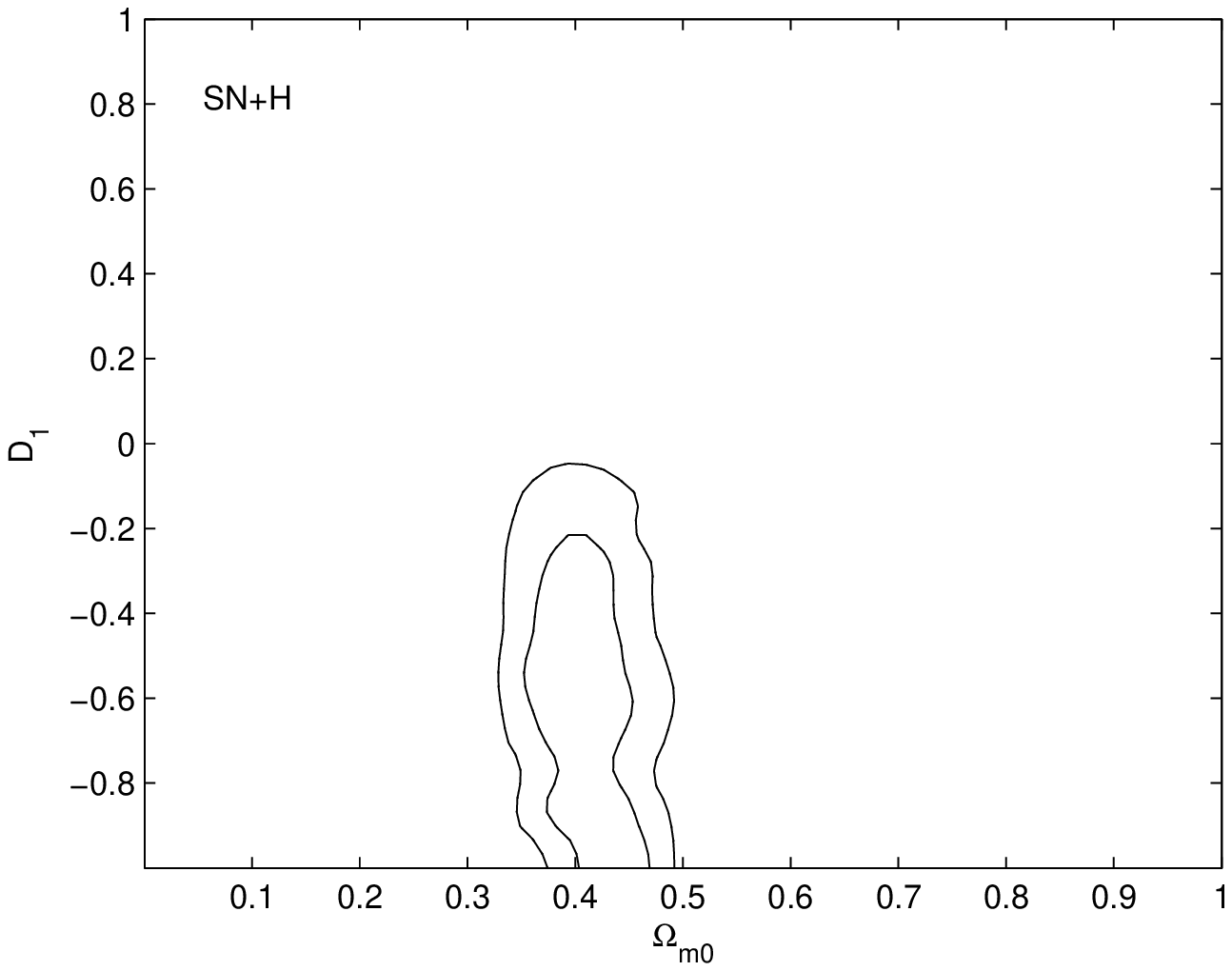}
\includegraphics[scale=0.5]{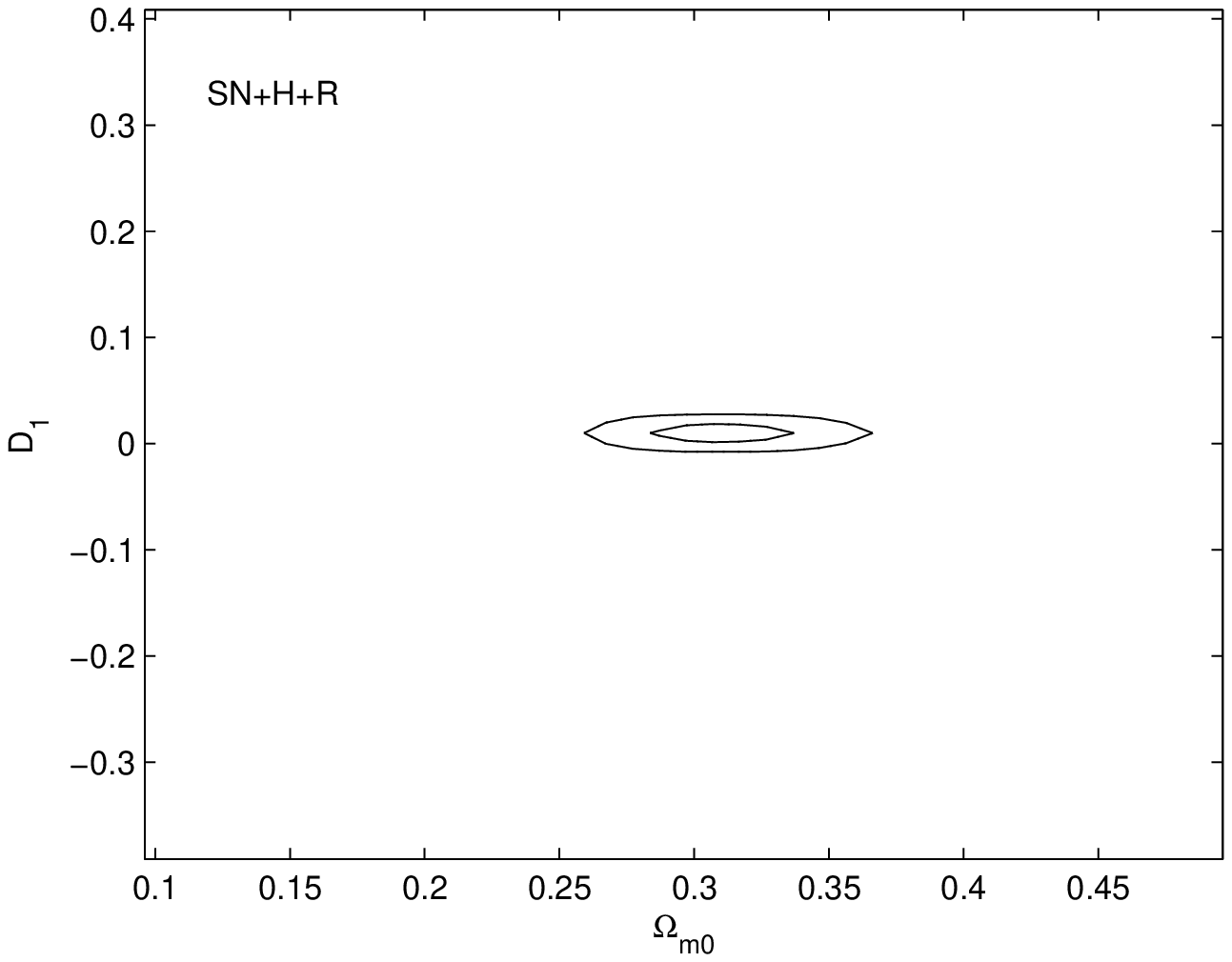}
\includegraphics[scale=0.5]{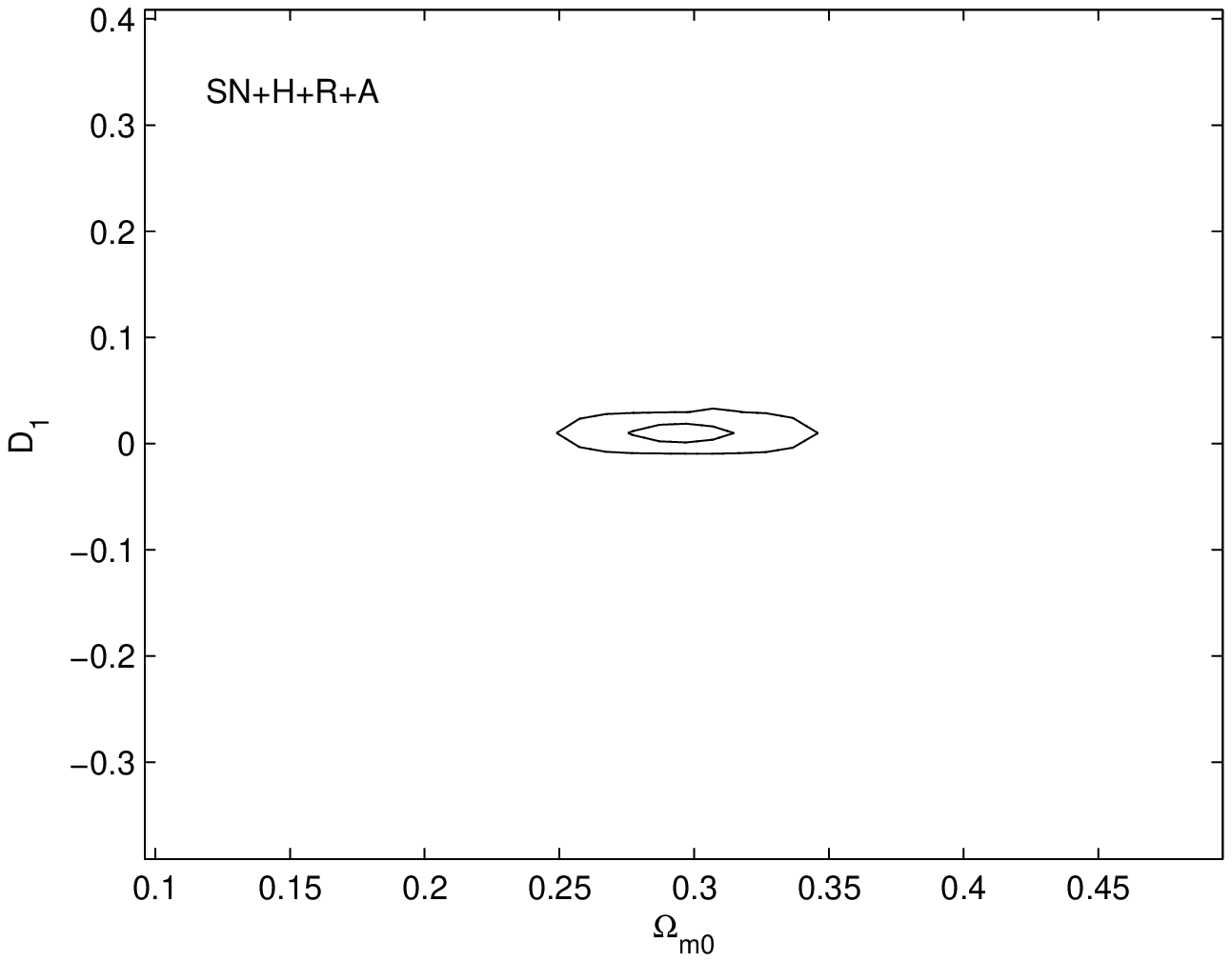}
\caption{Contour plots representing the $68\%$ and $95\%$ credible interval of the joint posterior probability distribution for $\Omega_{m,0}$ and $D_1$ oscillating DE model parameters.}
\label{fig:2}
\end{figure}

\begin{figure}[h]
\includegraphics[scale=0.5]{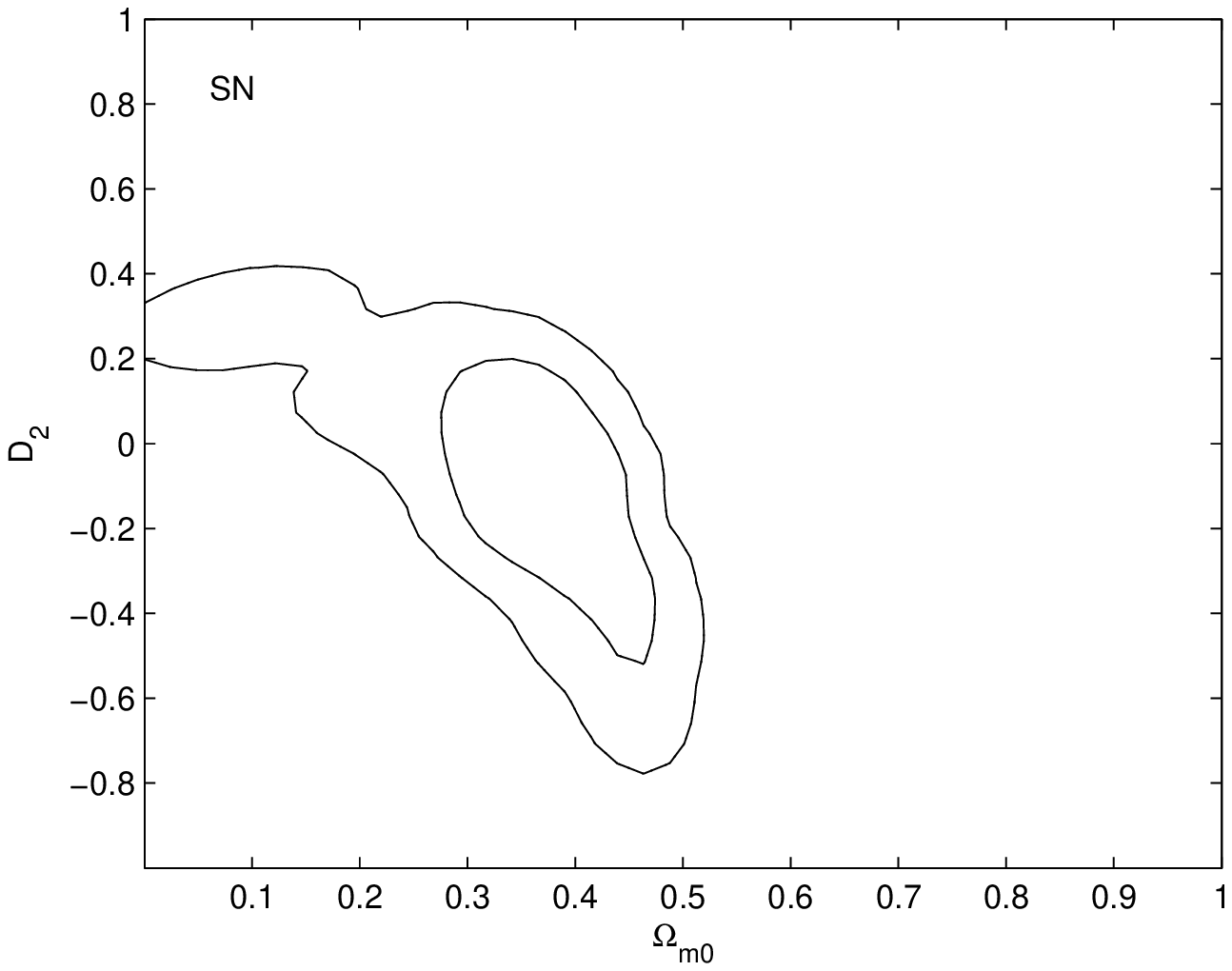}
\includegraphics[scale=0.5]{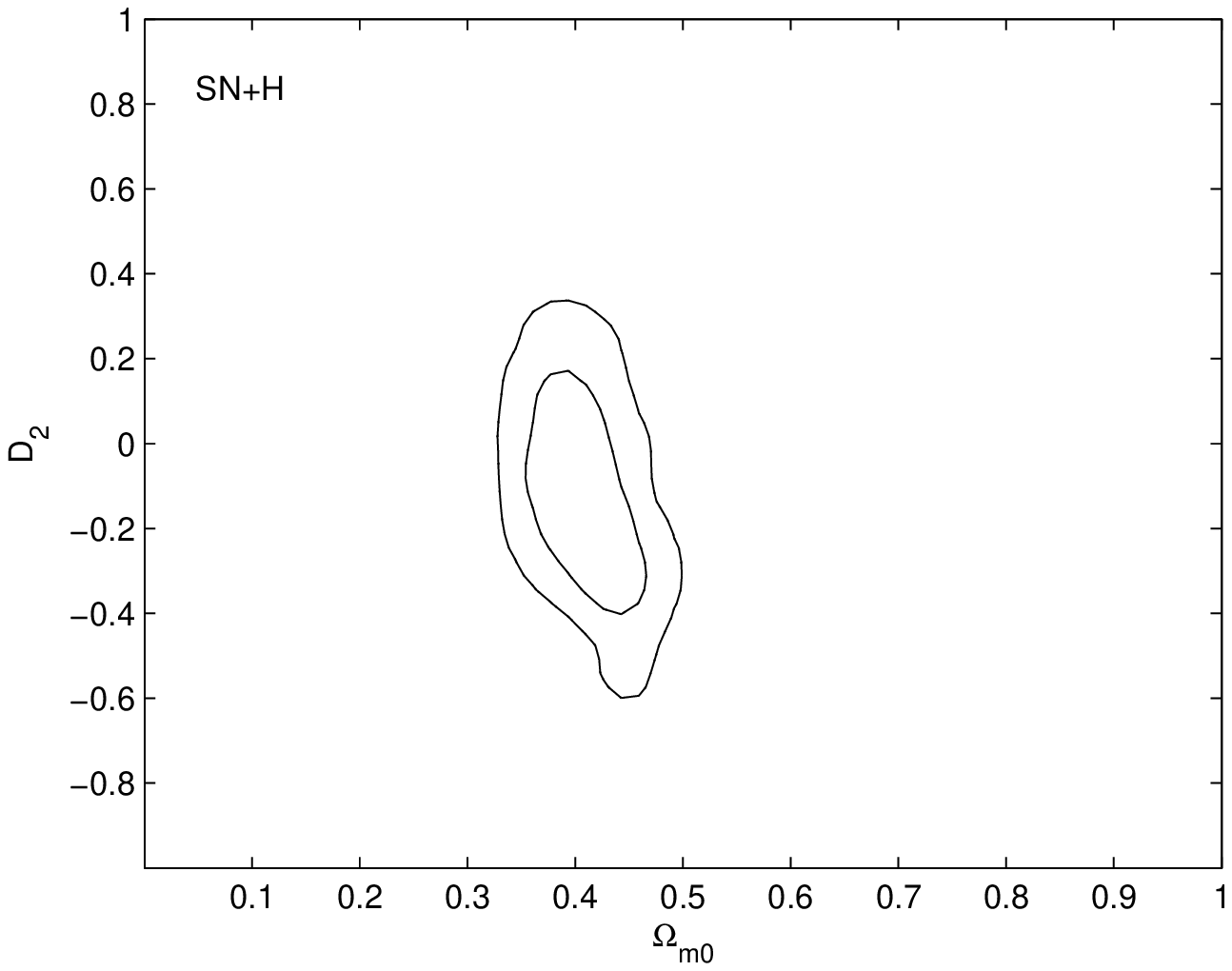}
\includegraphics[scale=0.5]{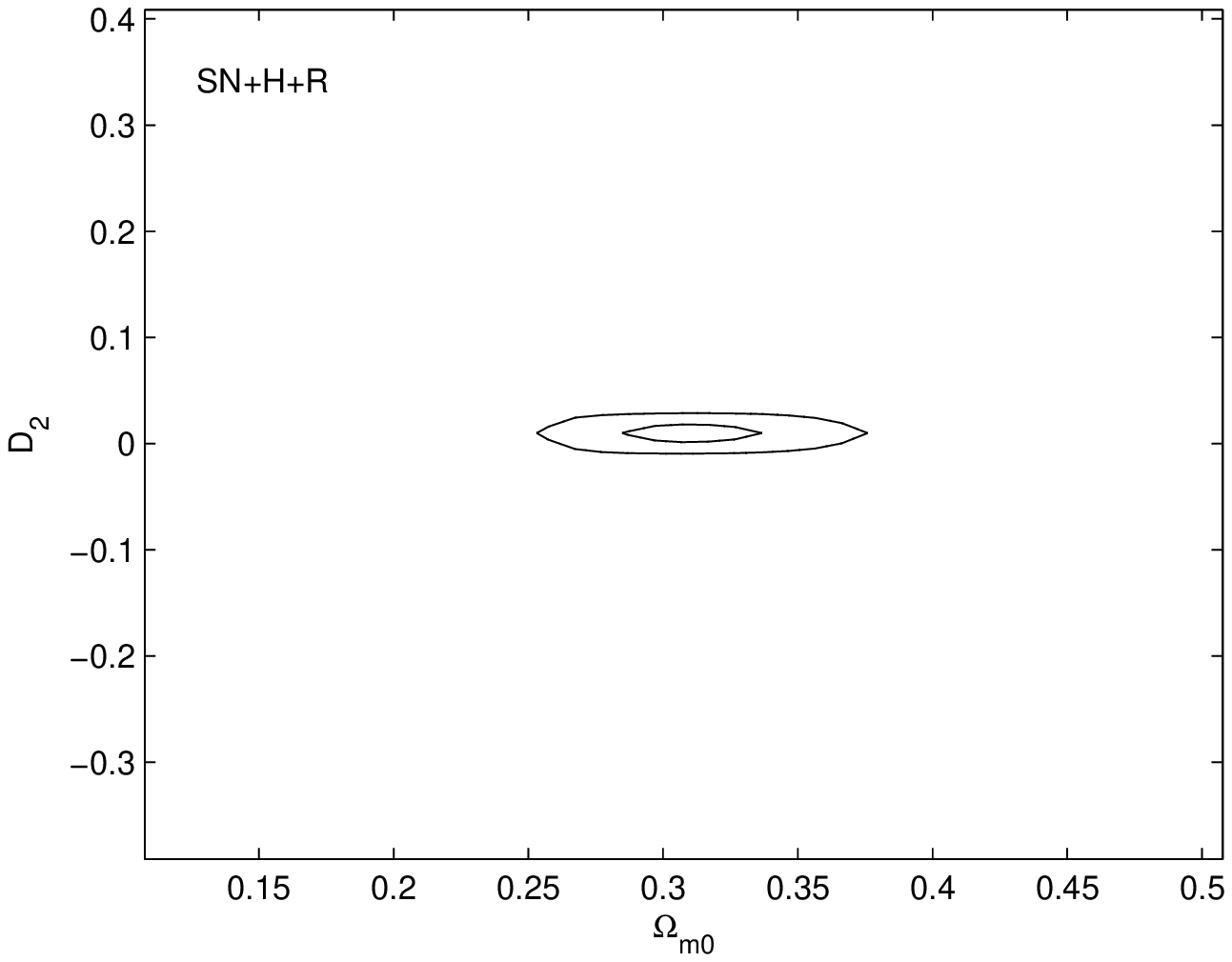}
\includegraphics[scale=0.5]{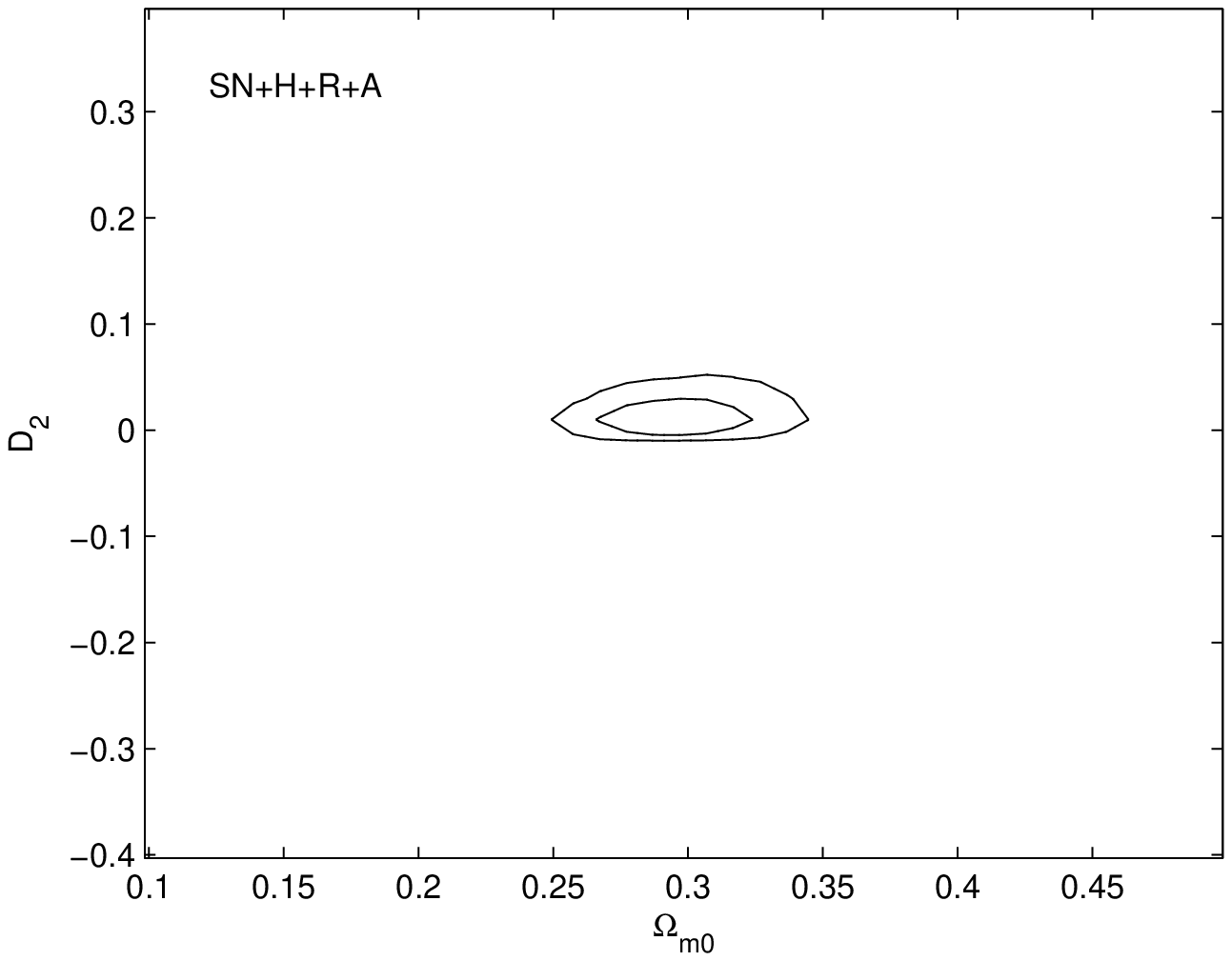}
\caption{Contour plots representing the $68\%$ and $95\%$ credible interval of the joint posterior probability distribution for $\Omega_{m,0}$ and $D_2$ oscillating DE model parameters.}
\label{fig:3}
\end{figure}

\begin{figure}[h]
\includegraphics[scale=0.5]{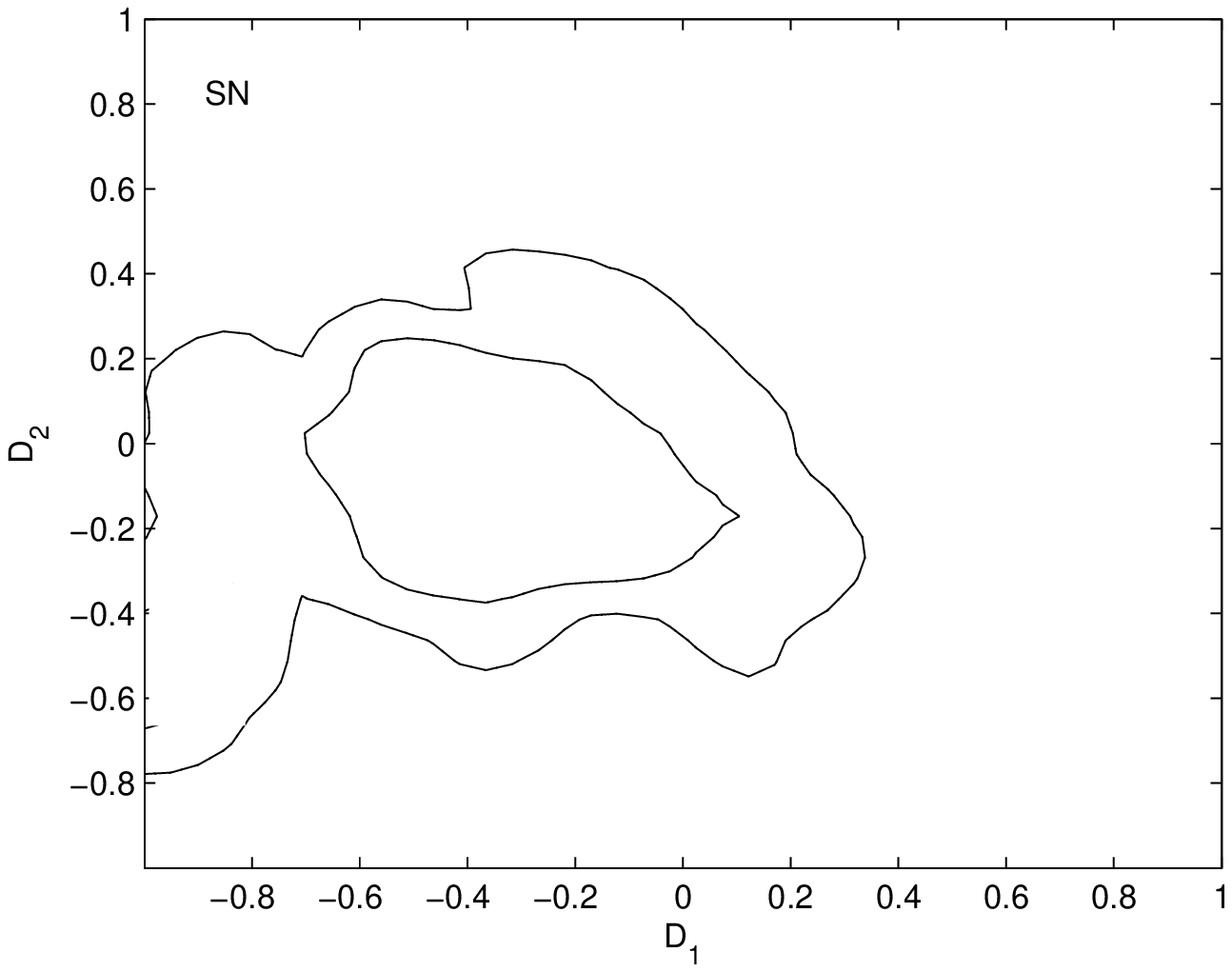}
\includegraphics[scale=0.5]{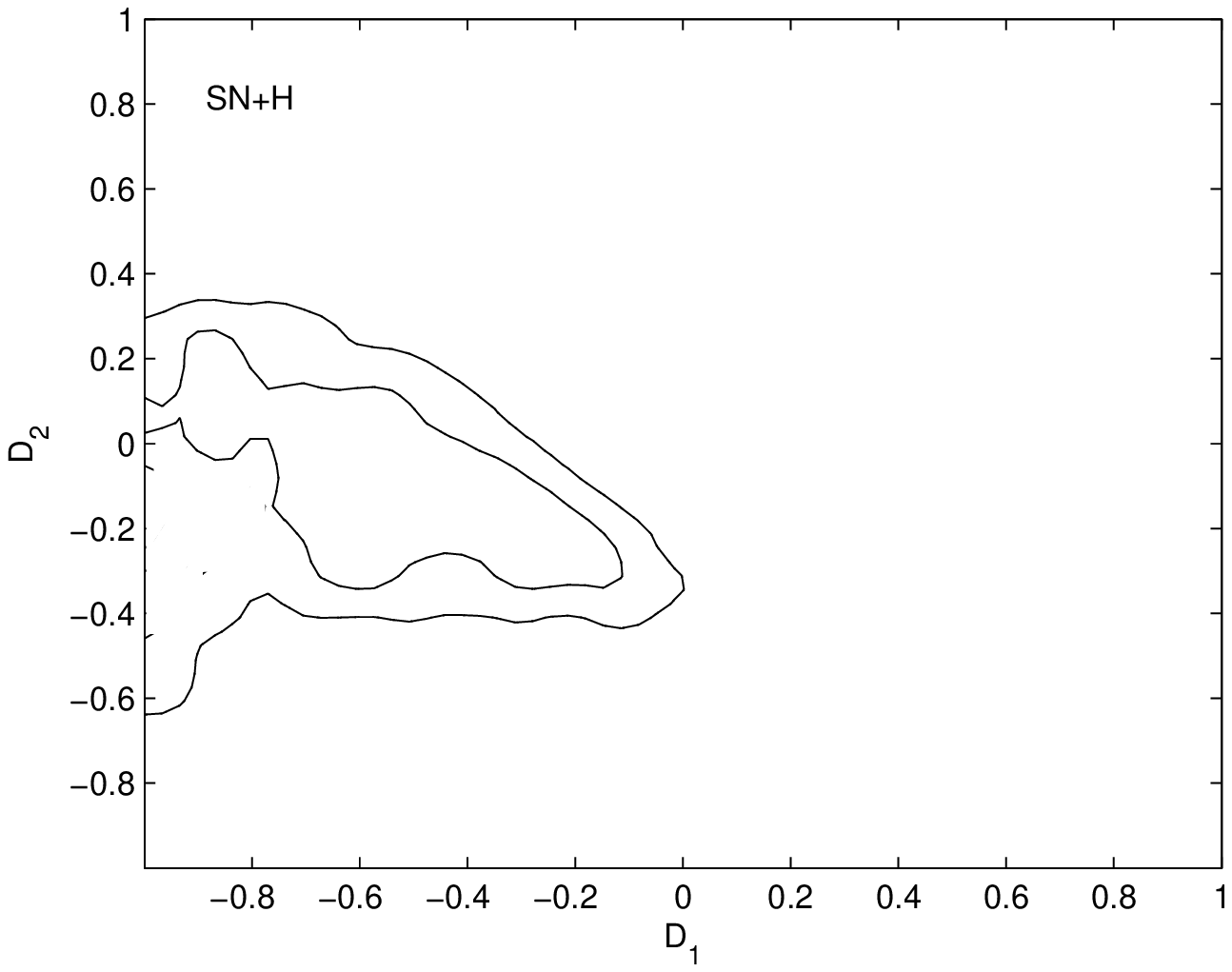}
\includegraphics[scale=0.5]{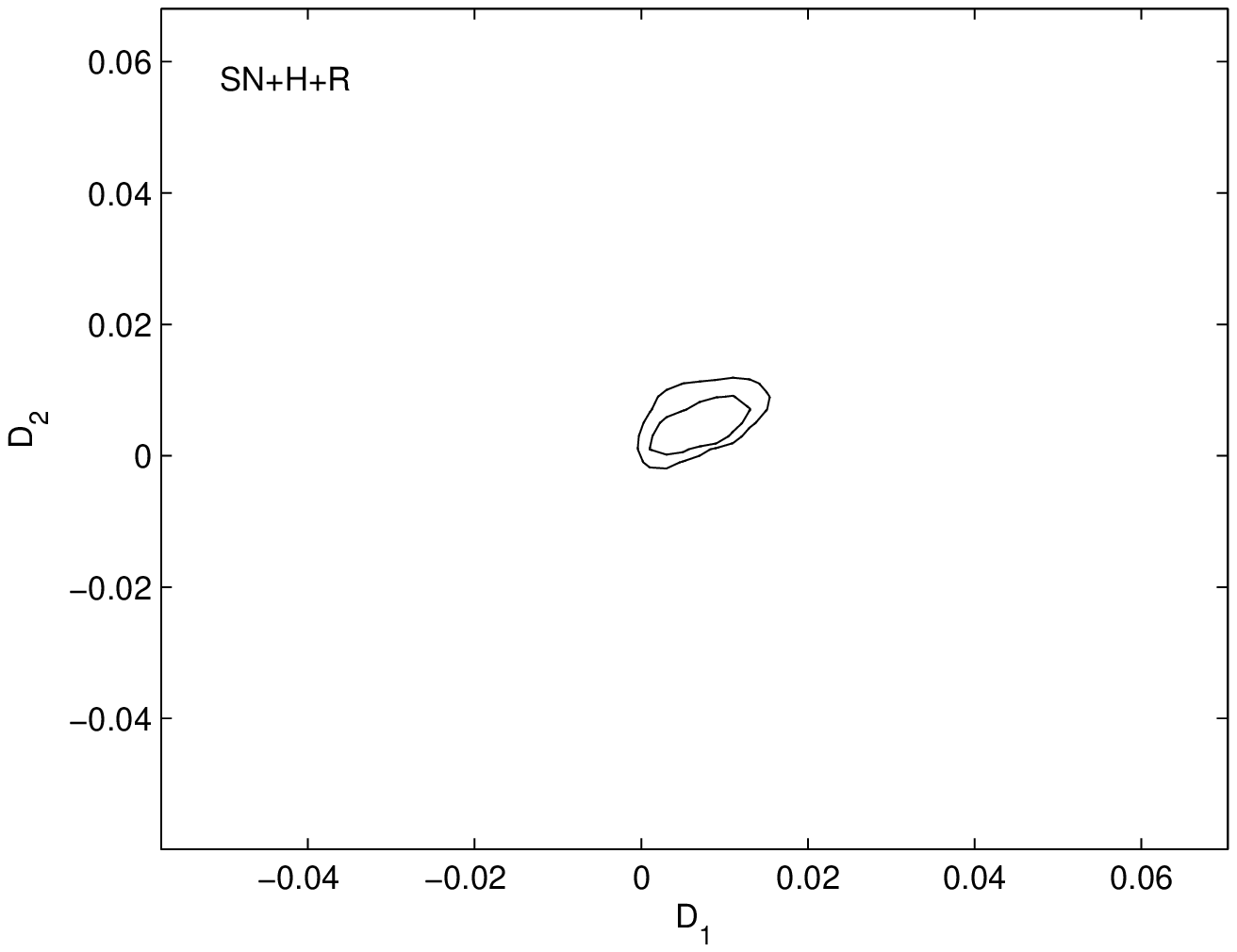}
\includegraphics[scale=0.5]{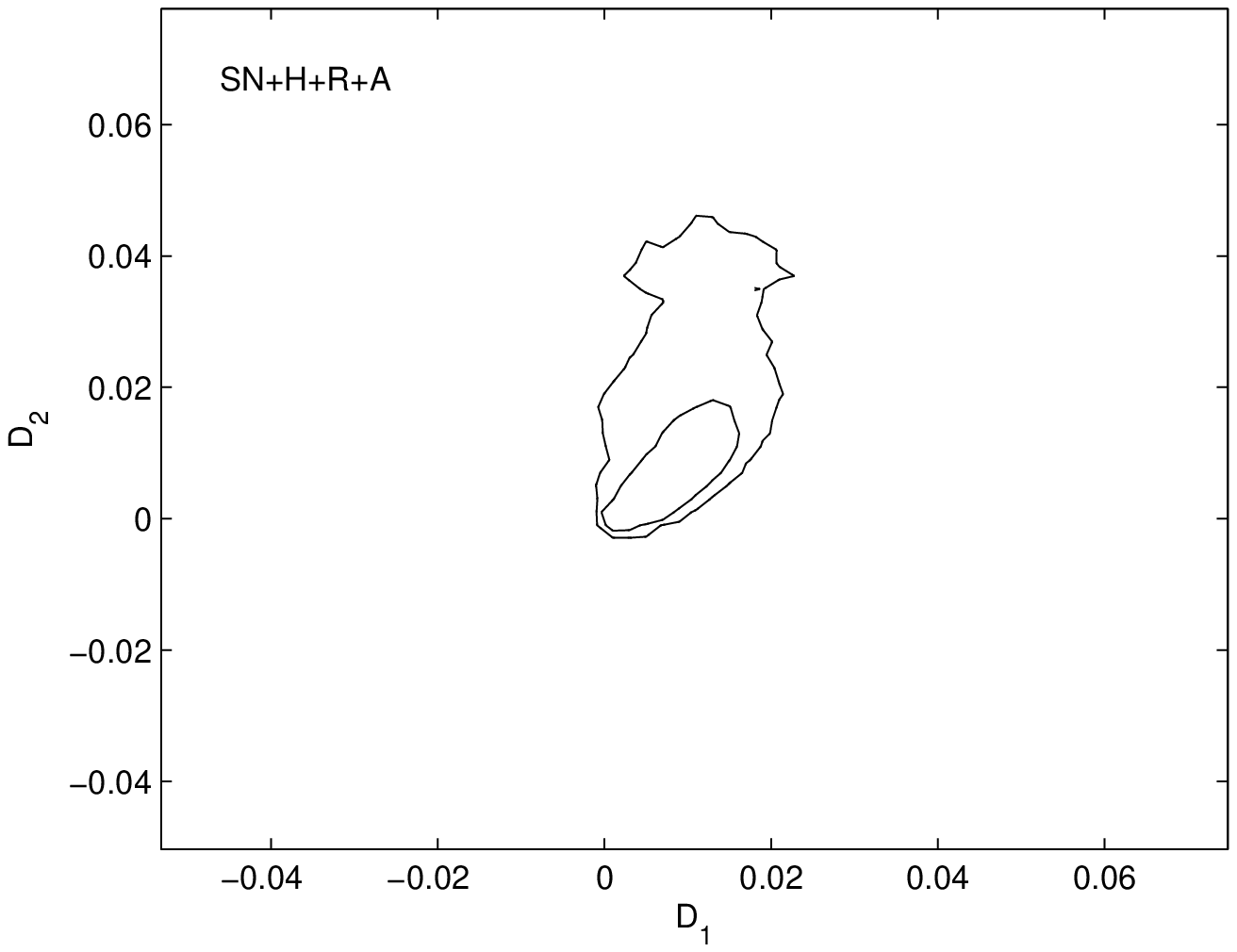}
\caption{Contour plots representing the $68\%$ and $95\%$ credible interval of the joint posterior probability distribution for $D_1$ and $D_2$ oscillating DE model parameters.}
\label{fig:4}
\end{figure}

The best fit values for model parameters (the mode of the posterior probability) are the same as the best fit values obtained by $\chi ^2$ minimization within the interval for parameters assumed before. Results, i.e. values for model parameters obtained via $\chi ^2$ minimization procedure are gathered in Table \ref{tab:1}.\\
Posterior probabilities for model parameters defined in the following way
\begin{eqnarray} 
P(\Omega_{m,0}|M,D)=\int \int P(\bar{\theta}|M,D) dD_1 dD_2 \nonumber \\
P(D_1|M,D)=\int \int P(\bar{\theta}|M,D) d\Omega_{m,0} dD_2 \nonumber \\
P(D_2|M,D)=\int \int P(\bar{\theta}|M,D) d\Omega_{m,0} dD_1 \nonumber \\
\end{eqnarray}
are presented on Figure \ref{fig:1}. The values of the mean for such distributions together with $68\%$ and $95\%$ credible interval are gathered in Table \ref{tab:1}. Two dimensional contour plots representing the $68\%$ and $95\%$ credible interval of the joint posterior probability distributions i.e. $P(\Omega_{m,0},D_1|M,D)=\int P(\bar{\theta}|M,D) dD_2$, $P(\Omega_{m,0},D_2|M,D)=\int P(\bar{\theta}|M,D) dD_1$, $P(D_1,D_2|M,D)= \int P(\bar{\theta}|M,D) d\Omega_{m,0}$ are presented on Figure \ref{fig:2}, \ref{fig:3} and \ref{fig:4} respectively. \\

We add constraints coming from observational H(z) data (N=9) \cite{Simon:2004tf, Samushia:2005, Wei:2007}. This data based on the differential ages ($\frac{dt}{dz}$) of the passively evolving galaxies which allow to estimate the relation $H(z) \equiv \frac{\dot{a}}{a} =-\frac{1}{1+z} \frac{dz}{dt}$. The posterior probability for model parameters has the following form  
\begin{equation}
P(\bar{\theta}|M,D) \propto \int \big ( \pi(\bar{\theta}|M) \exp \left[-0.5(\chi ^2_{SN}(\bar{\theta})+\chi ^2_{H}(\bar{\theta}) )\right] \big ) dH_0,
\end{equation}
where $\chi ^2_{H}(\bar{\theta})= \sum_{i=1}^{N} \left( \frac{H(z_i) -H_i(z_i)}{\sigma_i ^2}\right)^2$.

We also used constraints coming from so called CMB R shift parameter. In this case the posterior probability for model parameters has the following form
\begin{equation}
P(\bar{\theta}|M,D) \propto \int \big ( \pi(\bar{\theta}|M) \exp \left[-0.5(\chi ^2_{SN}(\bar{\theta})+\chi ^2_{H}(\bar{\theta})+\chi ^2_{R}(\bar{\theta})) \right] \big ) dH_0,
\end{equation}
where $\chi ^2_{R}(\bar{\theta}) =\left( \frac{R^{obs}-R^{th}}{\sigma_R} \right)^2$ and $R^{th}=\sqrt{\Omega_{{m},0}}\int_{0}^{z_{dec}}\frac{H_0}{H(z)}dz$, $R^{{obs}}=1.70 \pm 0.03$ for $z_{\mathrm{dec}}=1089$ \cite{Wang:2006ts}.\\

Finally we add constraints coming from the SDSS luminous red galaxies measurement of $A$ parameter ($A^{obs}=0.469 \pm 0.017$ for $z_{A}=0.35$) \cite{Eisenstein:2005}, which is related to the baryon acoustic oscillation peak and defined in the following way
$A^{th}=\sqrt{\Omega_{m,0}} \left (\frac{H(z_A)}{H_{0}} \right ) ^{-\frac{1}{3}} \left [ \frac{1}{z_{A}} \int_{0}^{z_{A}}\frac {H_0}{H(z)}dz\right]^{\frac{2}{3}}$.\\
This parameter was derived with assumption that $w(z)$ is a constant. Due to
that using this value to constraints varying $w(z)$ lead to systematic errors in
the parameter constraints \cite{Dick:2006ev}. The posterior probability has the following form
\begin{equation}
P(\bar{\theta}|M,D) \propto \int \big ( \pi(\bar{\theta}|M) \exp \left[-0.5(\chi ^2_{SN}(\bar{\theta})+\chi ^2_{H}(\bar{\theta})+\chi ^2_{R}(\bar{\theta})+\chi ^2_{A}(\bar{\theta})) \right] \big ) dH_0,
\end{equation}
where $\chi ^2_{A}(\bar{\theta})=\left( \frac{A^{th}-A^{obs}}{\sigma_{A}}\right)^2$.\\
 
\begin{table}[h]
\centering
\begin{tabular}{c||cccc||cccc|}
\hline
& SN & & & & SN+H & & &  \\
\hline
&Best fit&Mean&$68\%$&$95\%$&Best fit&Mean&$68\%$&$95\%$\\
\hline
$\Omega_{m,0}$&$0.16$&$0.35$&$<0.30,0.43>$&$<0.09,0.47>$&$0.41$&$0.41$&$<0.38,0.44>$&$<0.35,0.47>$\\
$D_1$&$0.17$&$-0.37$&$<-0.67,-0.10>$&$<-0.95,0.19>$&$-0.99$&$-0.58$&$<-0.86,-0.32>$&$<-0.98,-0.13>$\\
$D_2$&$-0.004$&$-0.10$&$<-0.31,0.11>$&$<-0.59,0.32>$&$0.16$&$-0.12$&$<-0.30,0.07>$&$<-0.46,0.25>$\\
\hline
$\chi^2$&$194.35$&&&&$206.23$&&&\\
&&&&&&&&\\
\hline
\hline
& SN+H+R & & & & SN+H+R+A & & &\\
\hline
&Best fit&Mean&$68\%$&$95\%$&Best fit&Mean&$68\%$&$95\%$\\
\hline
$\Omega_{m,0}$&$0.31$&$0.31$&$<0.29,0.34>$&$<0.26,0.36>$&$0.30$&$0.30$&$<0.28,0.31>$&$<0.26,0.33>$\\
$D_1$&$0.006$&$0.007$&$<0.003,0.01>$&$<0.001,0.014>$&$0.007$&$0.009$&$<0.004,0.015>$&$<0.001,0.021>$\\
$D_2$&$0.001$&$0.005$&$<0.002,0.008>$&$<-0.0001,0.011>$&$0.003$&$0.013$&$<0.002,0.02>$&$<-0.00003,0.04>$\\
\hline
$\chi^2$&$210.95$&&&&$212.01$&&&\\
&&&&&&&&\\
\hline
\end{tabular}
\caption{Values for oscillating DE model parameters obtained via $\chi^2$ minimization (best fit), values of the mean with the $68\%$ and $95\%$ credible intervals obtained from the posterior probability distribution for considered oscillating DE model parameter. }
\label{tab:1}
\end{table}

As one can see after inclusion all data to the analysis we obtain the values for $D_1$ and $D_2$ parameters which are close to zero. Due to that we also consider two models which are special cases of the oscillating dark energy model:

\begin{enumerate}
    \item Osc DE 1: $C_2=0 \Rightarrow w_X(z)=-1+C_1(1+z)^3\cos(\ln (1+z))$, \\
     $H(z)=H_0 \sqrt{\Omega_{\Lambda,0}\exp(-0.9C_1)\exp{\bigg(0.3C_1(1+z)^{3}\Big[3\cos(\ln(1+z))+\sin(\ln(1+z))\Big]\bigg)}
+\Omega_{m,0}(1+z)^{3} + \Omega_{r,0}(1+z)^{4}}$, where $\Omega_{r,0}\simeq 0.5 * 10^{-4}$ and $\Omega_{\Lambda,0}=1-\Omega_{m,0}-\Omega_{r,0}$.
    \item Osc DE 2: $C_1=0 \Rightarrow w_X(z)=-1+C_2(1+z)^3\sin(\ln (1+z))$, \\
     $H(z)=H_0 \sqrt{\Omega_{\Lambda,0}\exp(0.3C_2)\exp{\bigg(-0.3C_2(1+z)^{3}\Big[\cos(\ln(1+z))-3\sin(\ln(1+z))\Big]\bigg)}
+\Omega_{m,0}(1+z)^{3} + \Omega_{r,0}(1+z)^{4}}$, where $\Omega_{r,0}\simeq 0.5 * 10^{-4}$ and $\Omega_{\Lambda,0}=1-\Omega_{m,0}-\Omega_{r,0}$
\end{enumerate}

\begin{figure}[h]
\includegraphics[scale=0.5]{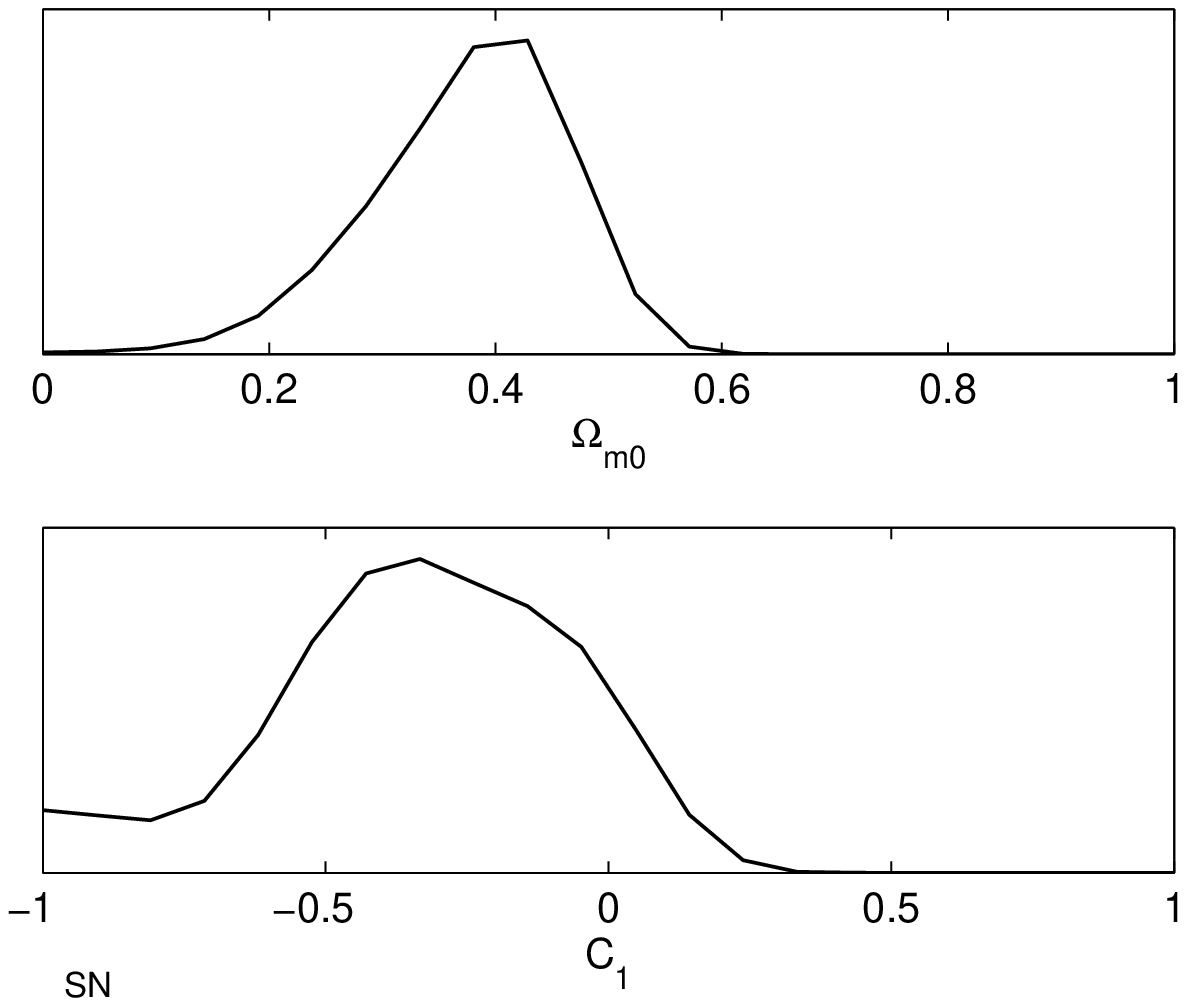}
\includegraphics[scale=0.5]{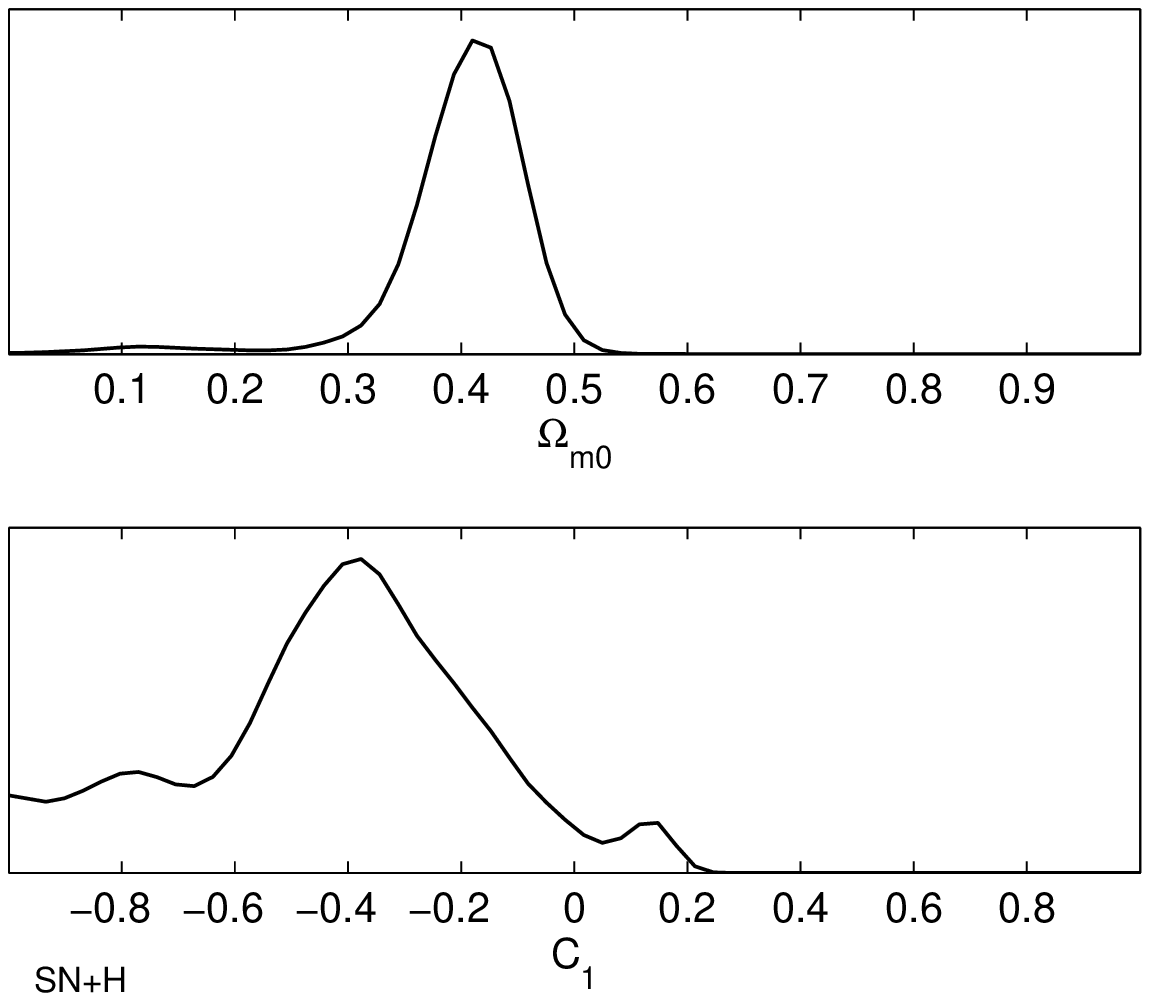}
\includegraphics[scale=0.5]{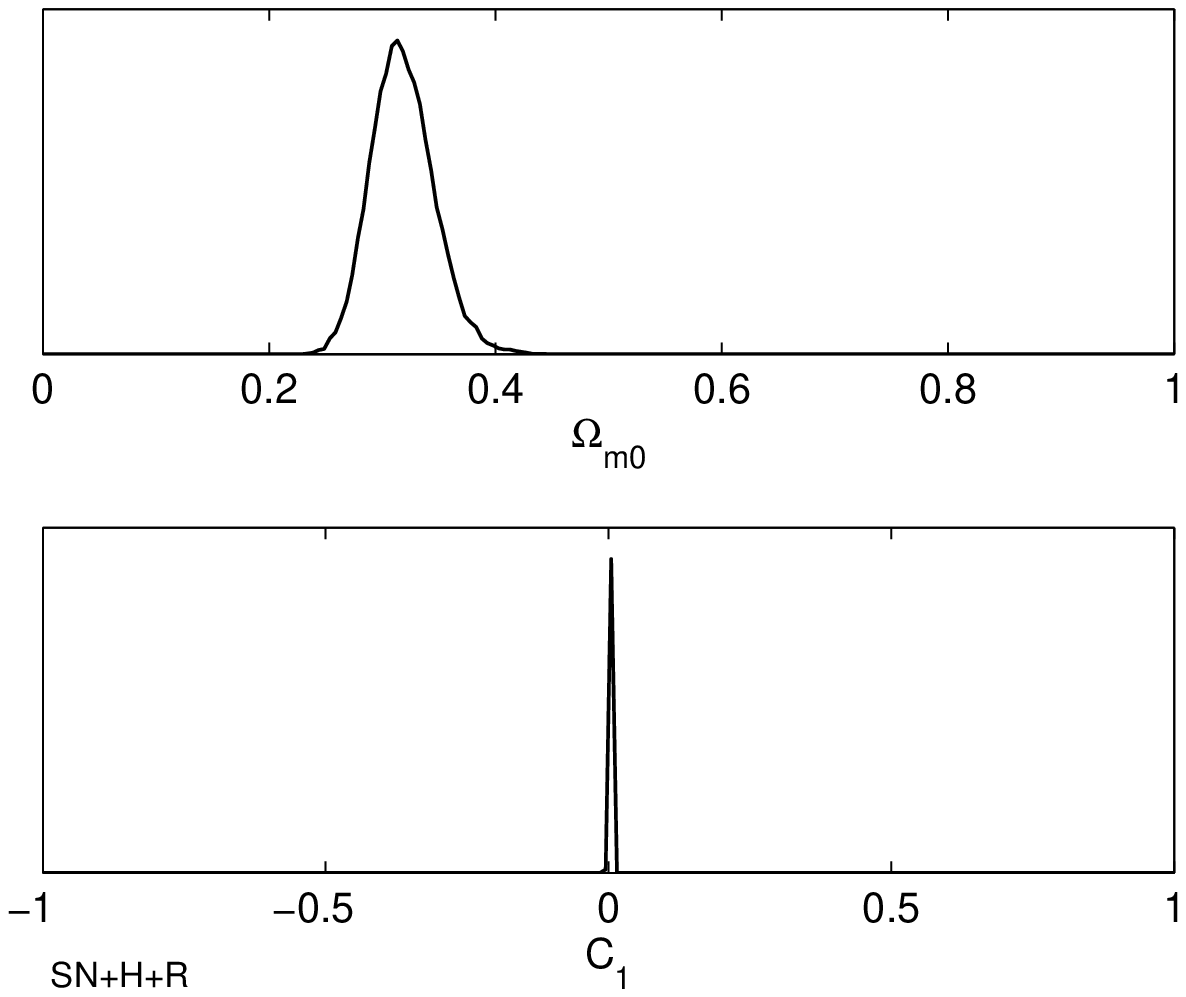}
\includegraphics[scale=0.5]{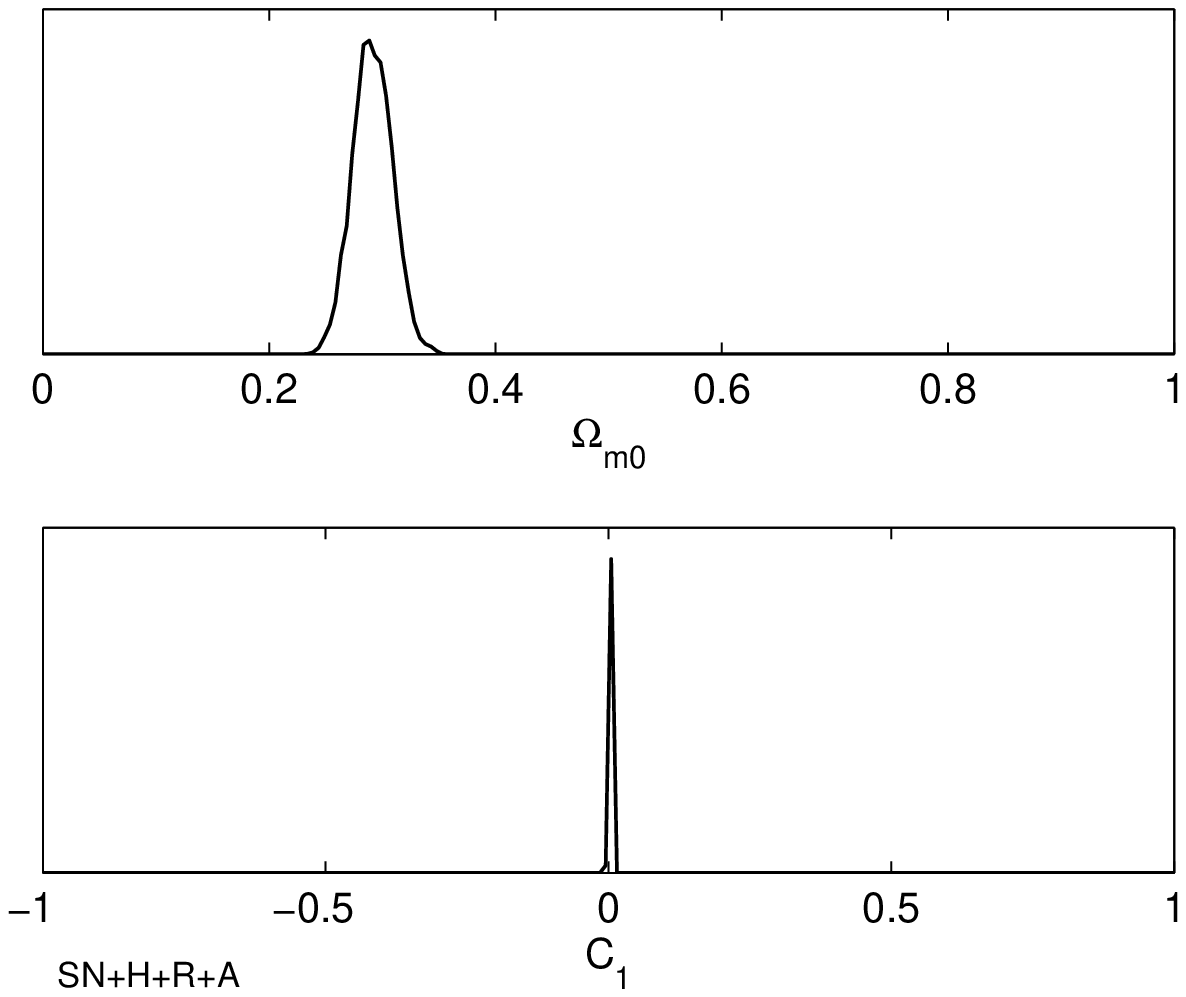}
\caption{Posterior probability distributions for Osc DE 1 model parameters.} 
\label{fig:5}
\end{figure}

\begin{figure}[h]
\includegraphics[scale=0.5]{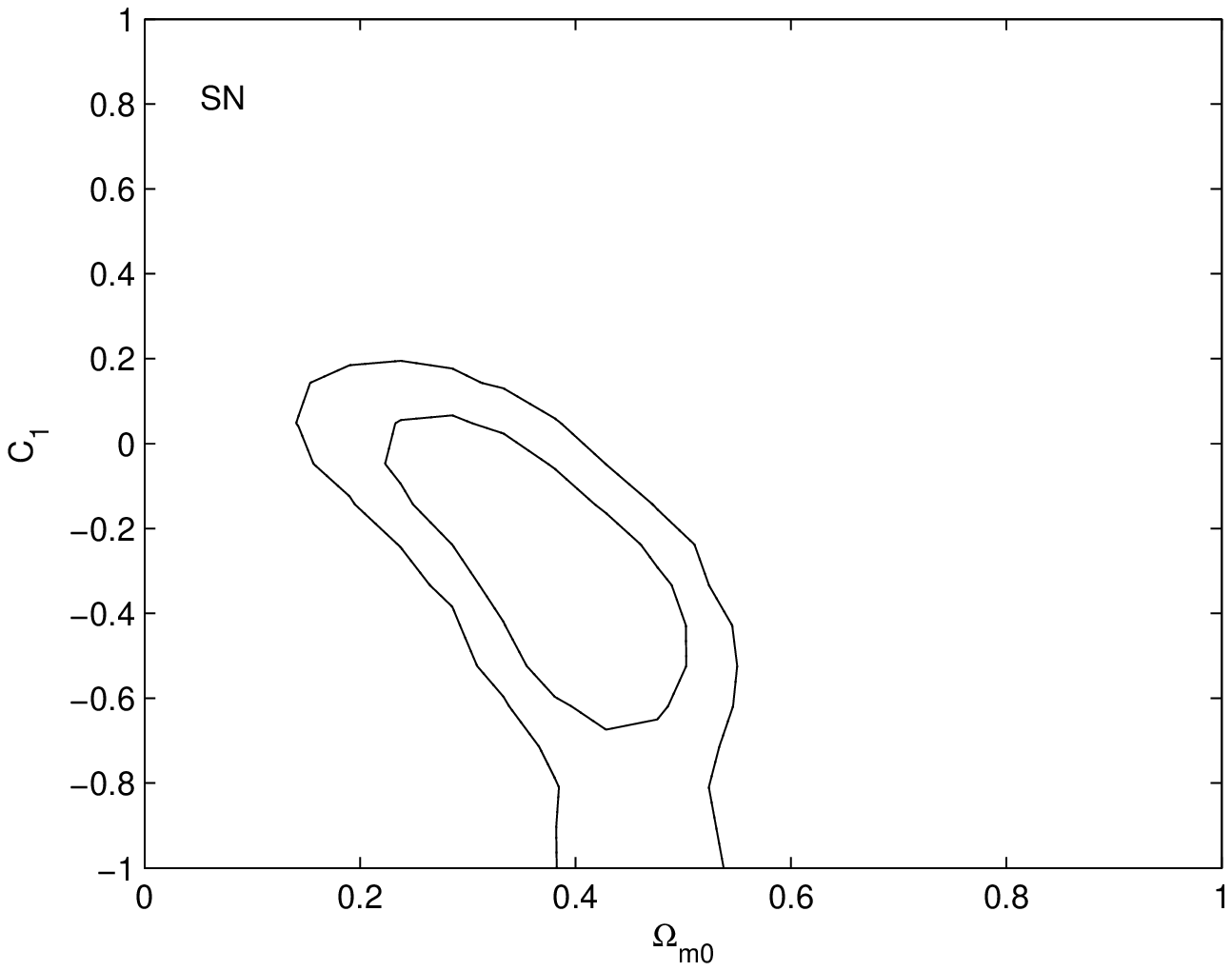}
\includegraphics[scale=0.5]{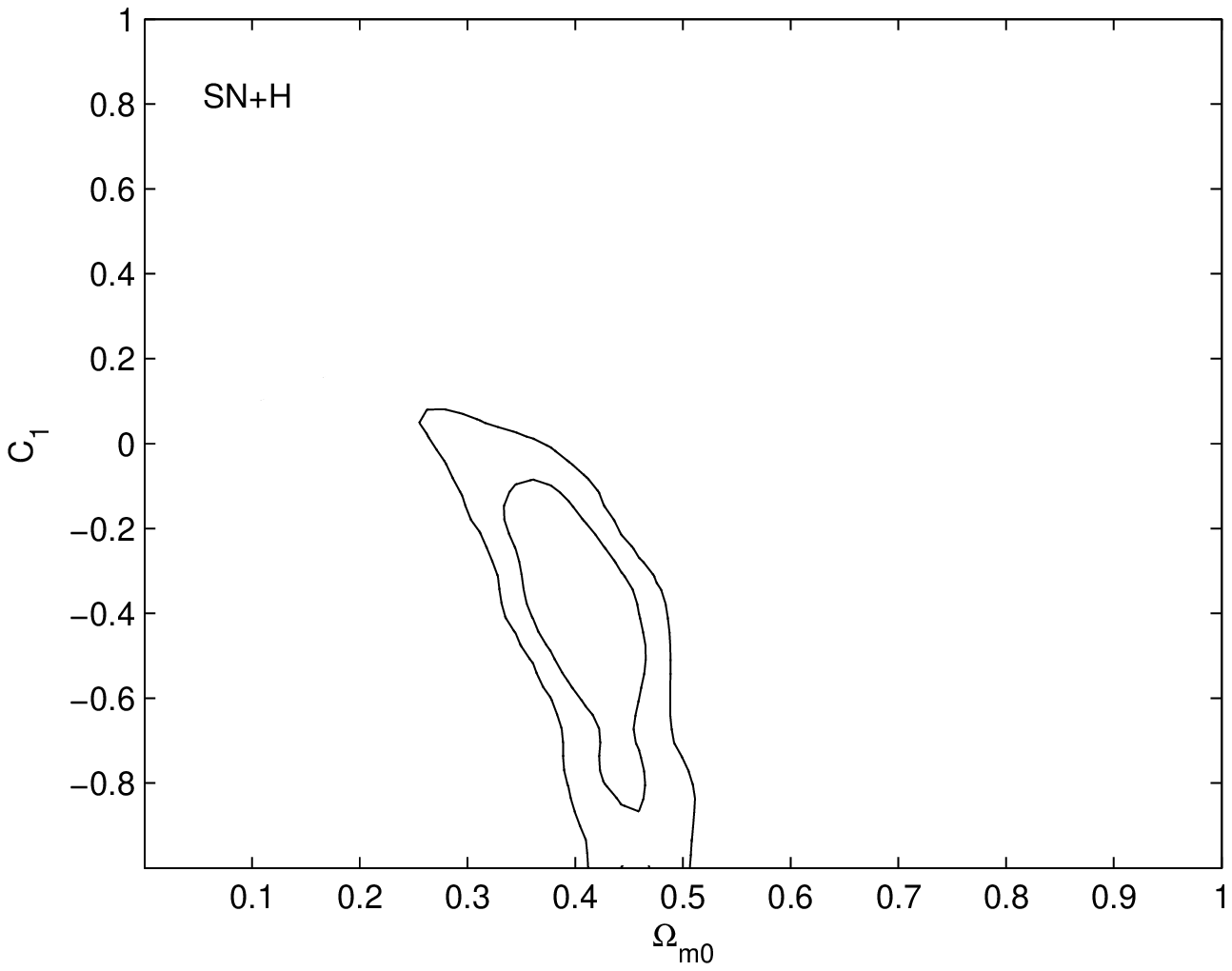}
\includegraphics[scale=0.5]{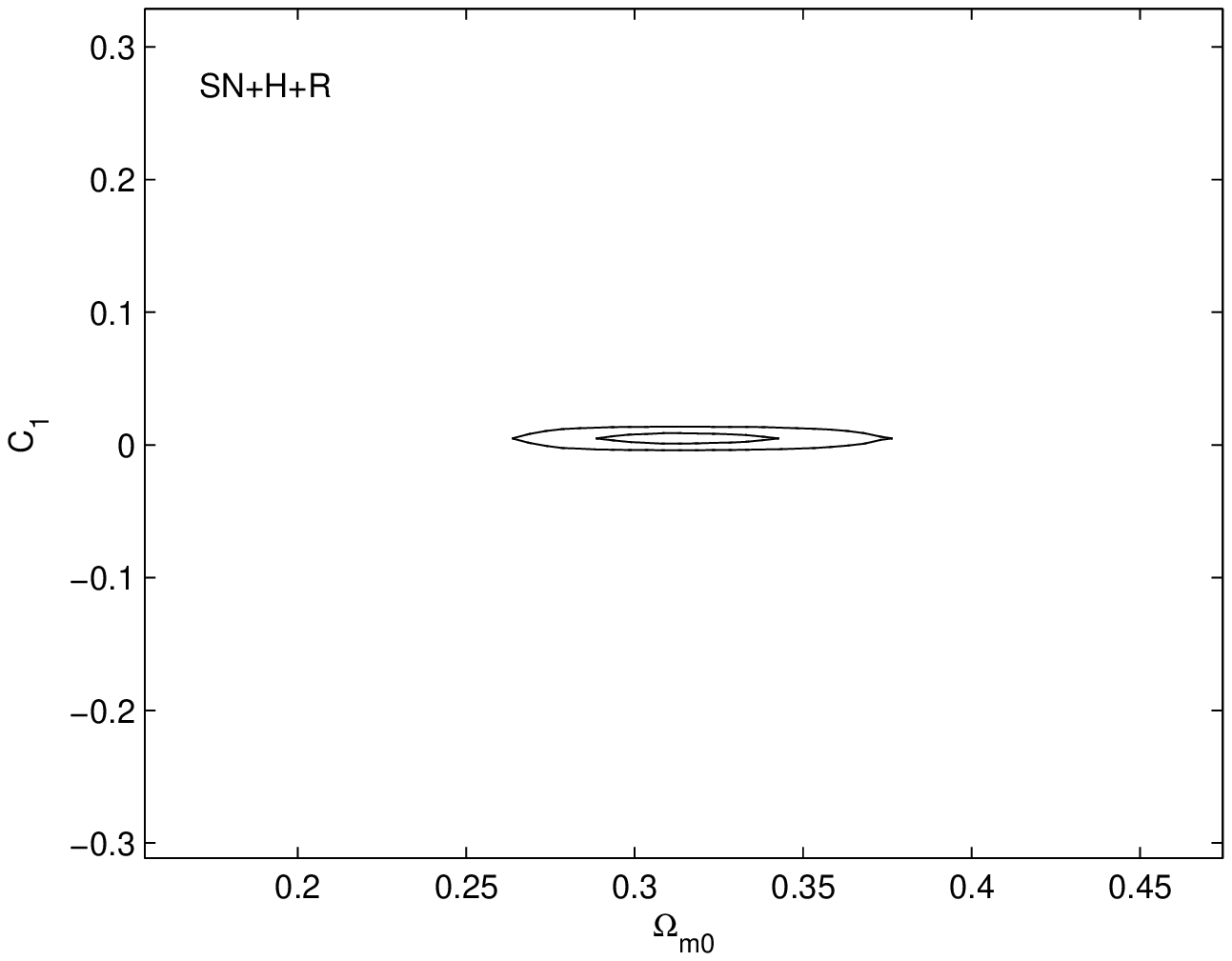}
\includegraphics[scale=0.5]{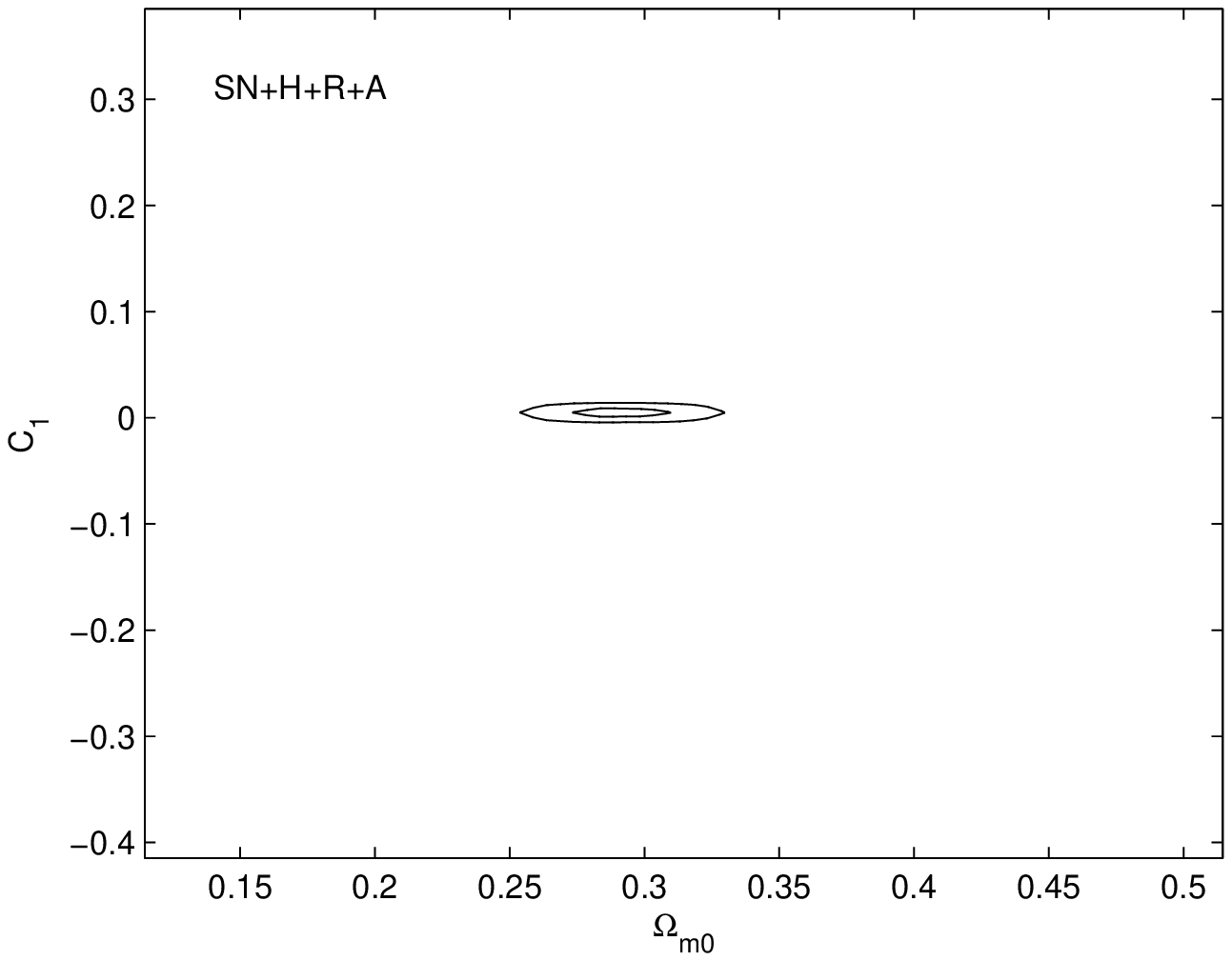}
\caption{Contour plots representing the $68\%$ and $95\%$ credible interval of the joint posterior probability distribution for $\Omega_{m,0}$ and $C_1$ Osc DE 1 model parameters.} 
\label{fig:6}
\end{figure}

\begin{figure}[h]
\includegraphics[scale=0.5]{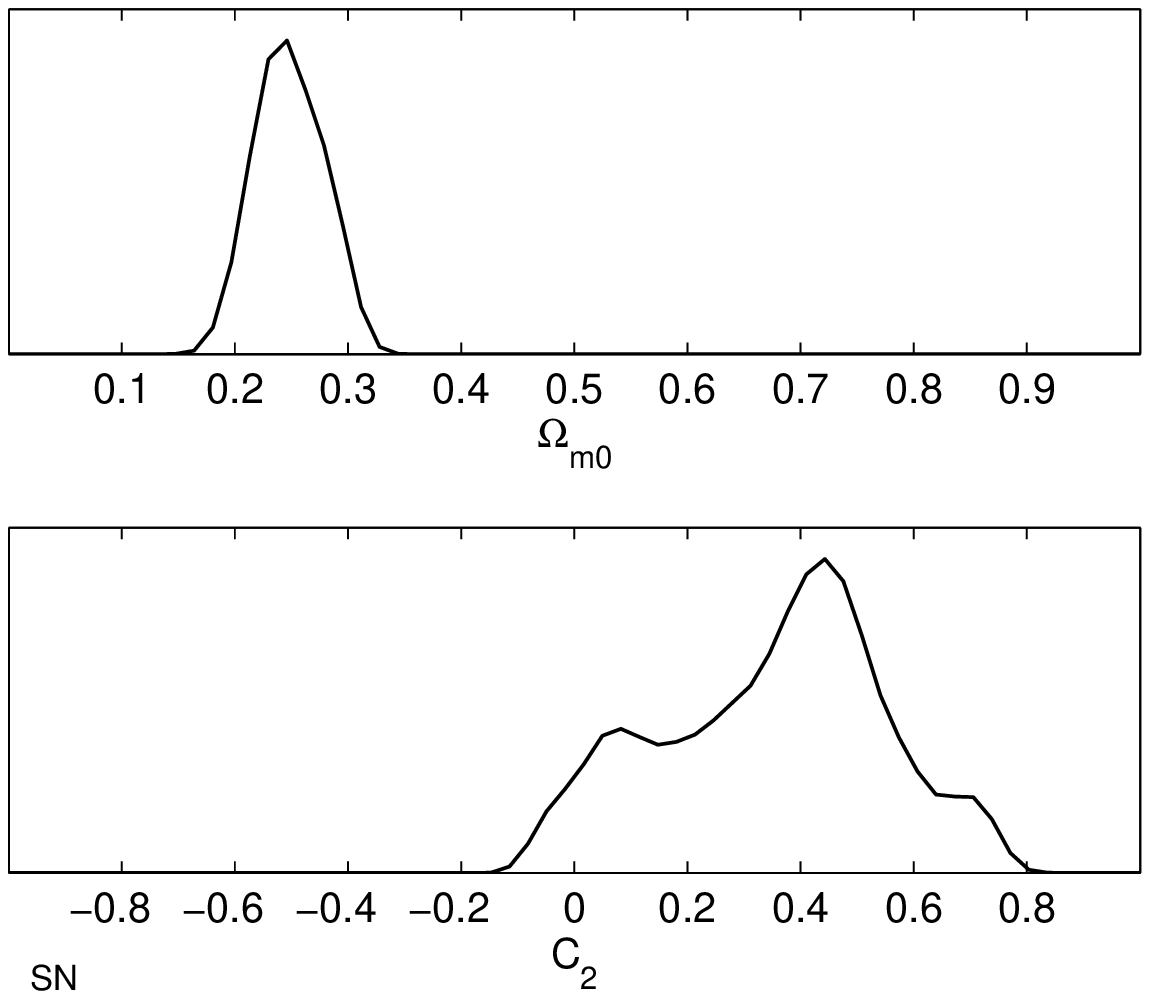}
\includegraphics[scale=0.5]{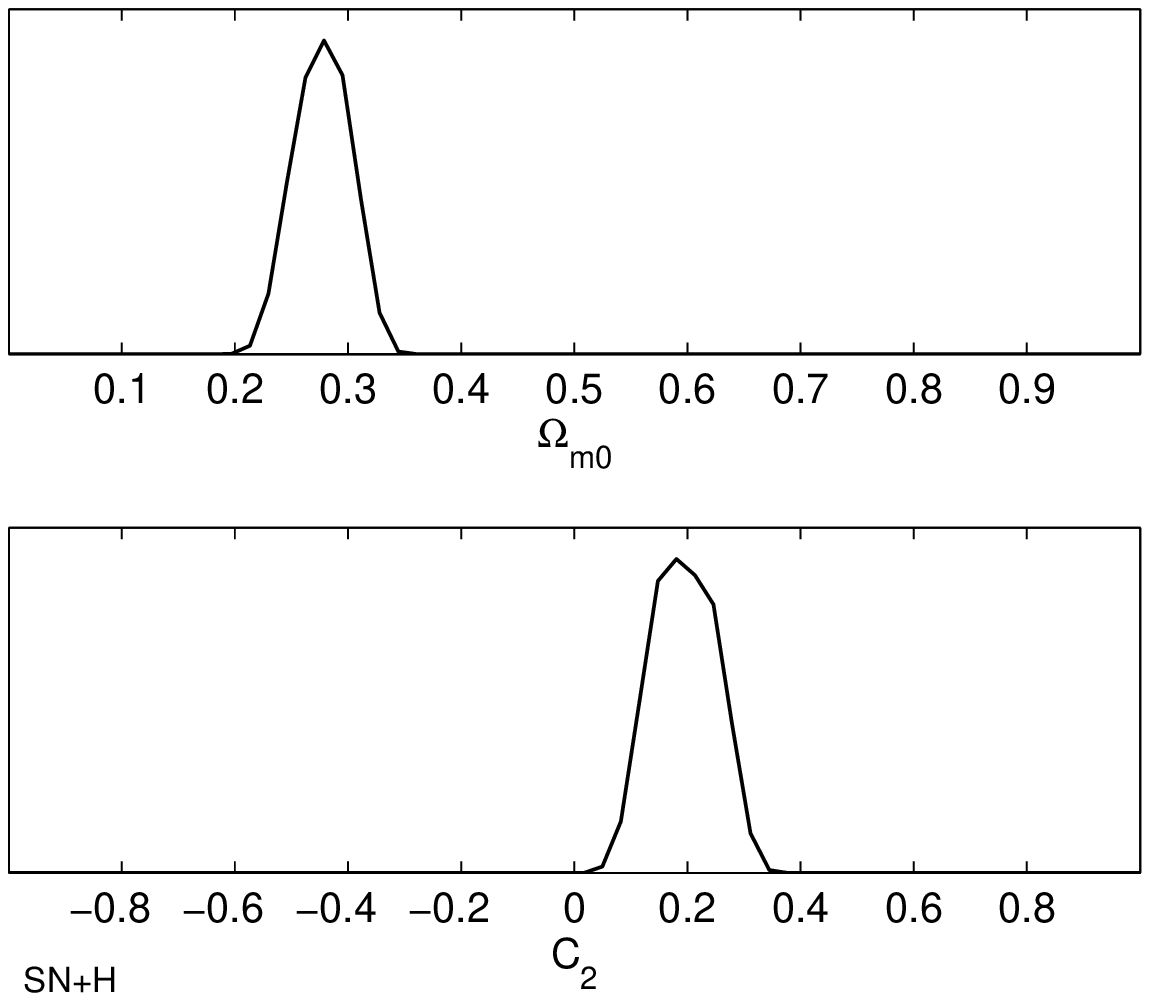}
\includegraphics[scale=0.5]{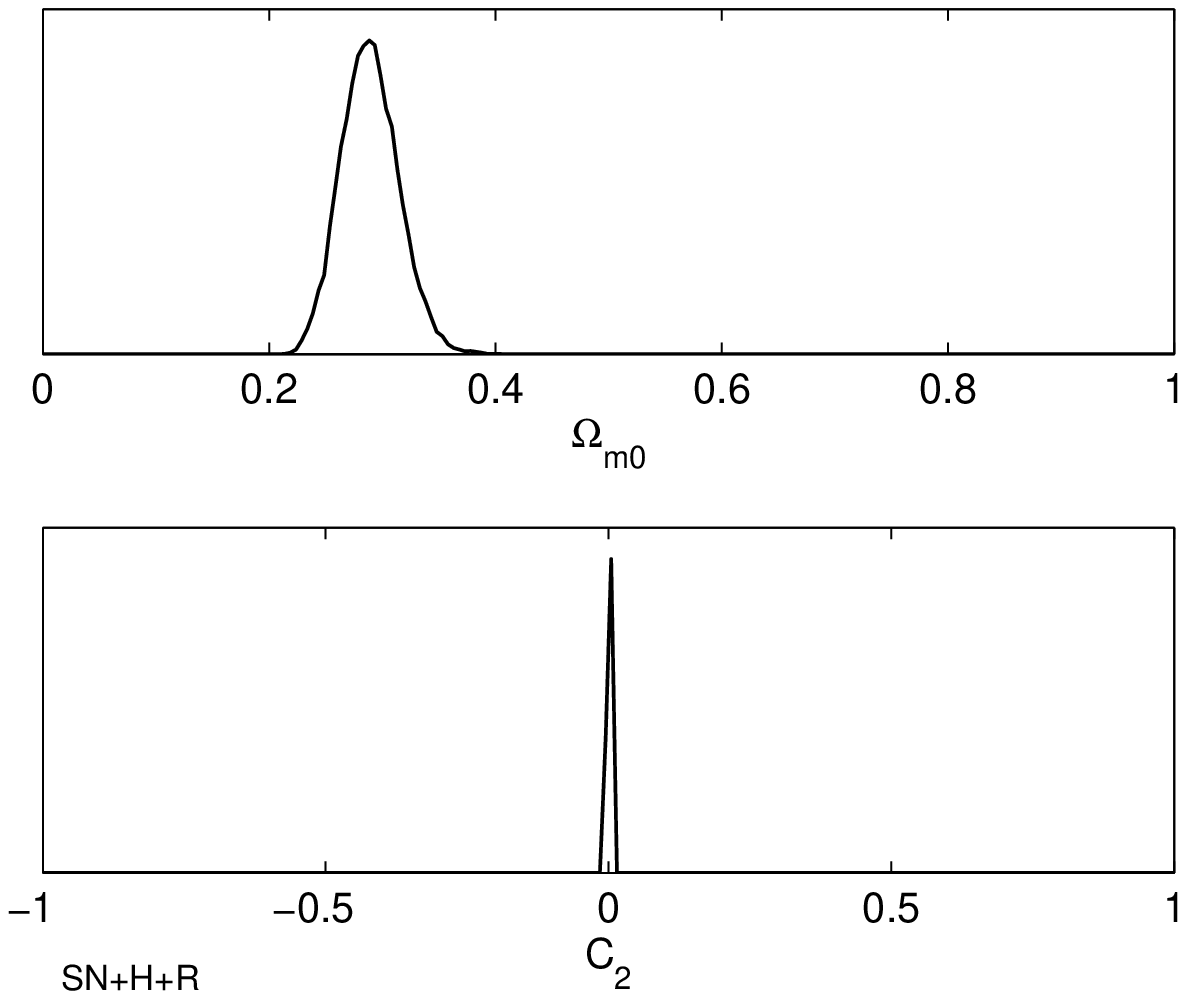}
\includegraphics[scale=0.5]{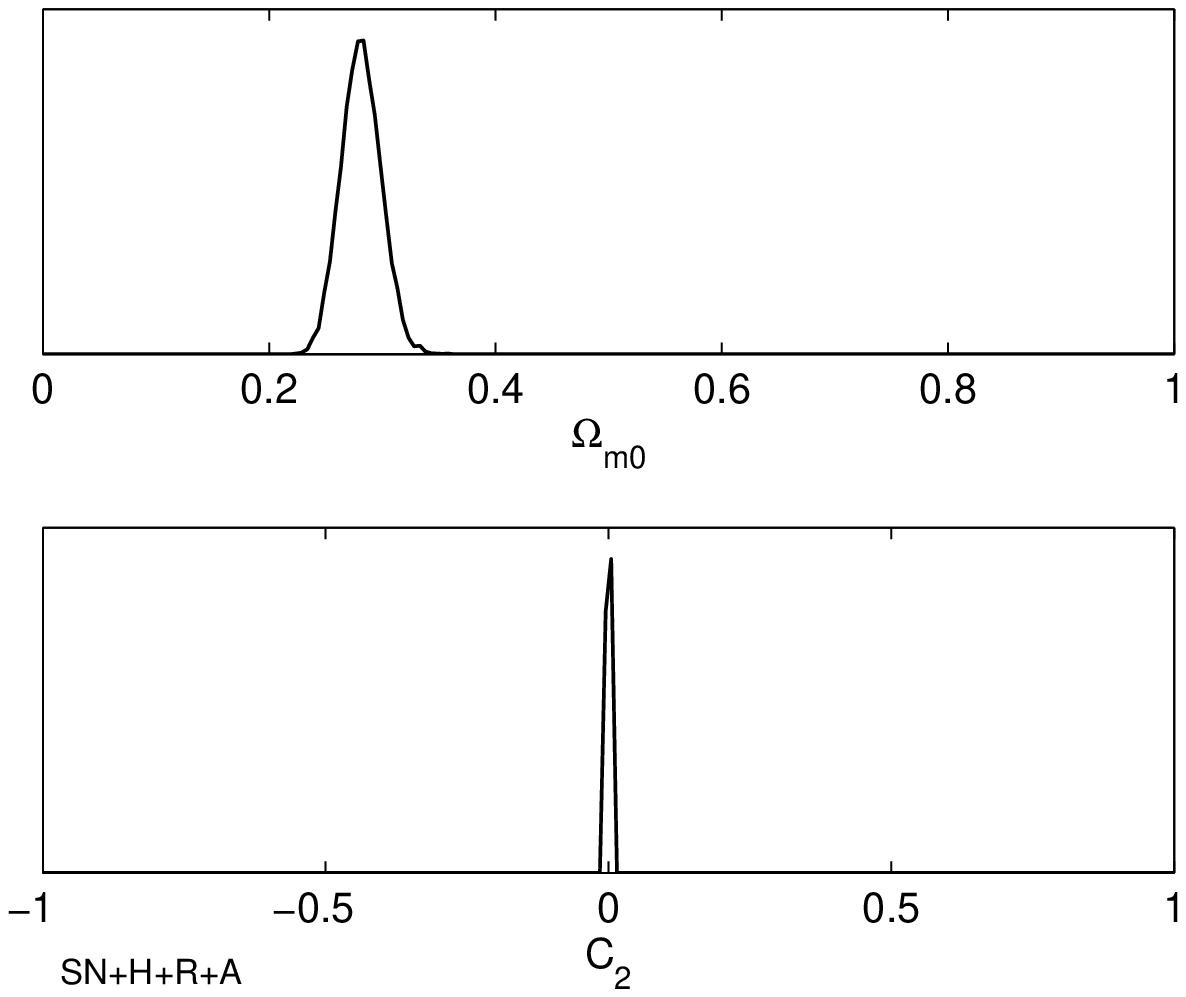}
\caption{Posterior probability distributions for Osc DE 2 model parameters.} 
\label{fig:7}
\end{figure}

\begin{figure}[h]
\includegraphics[scale=0.5]{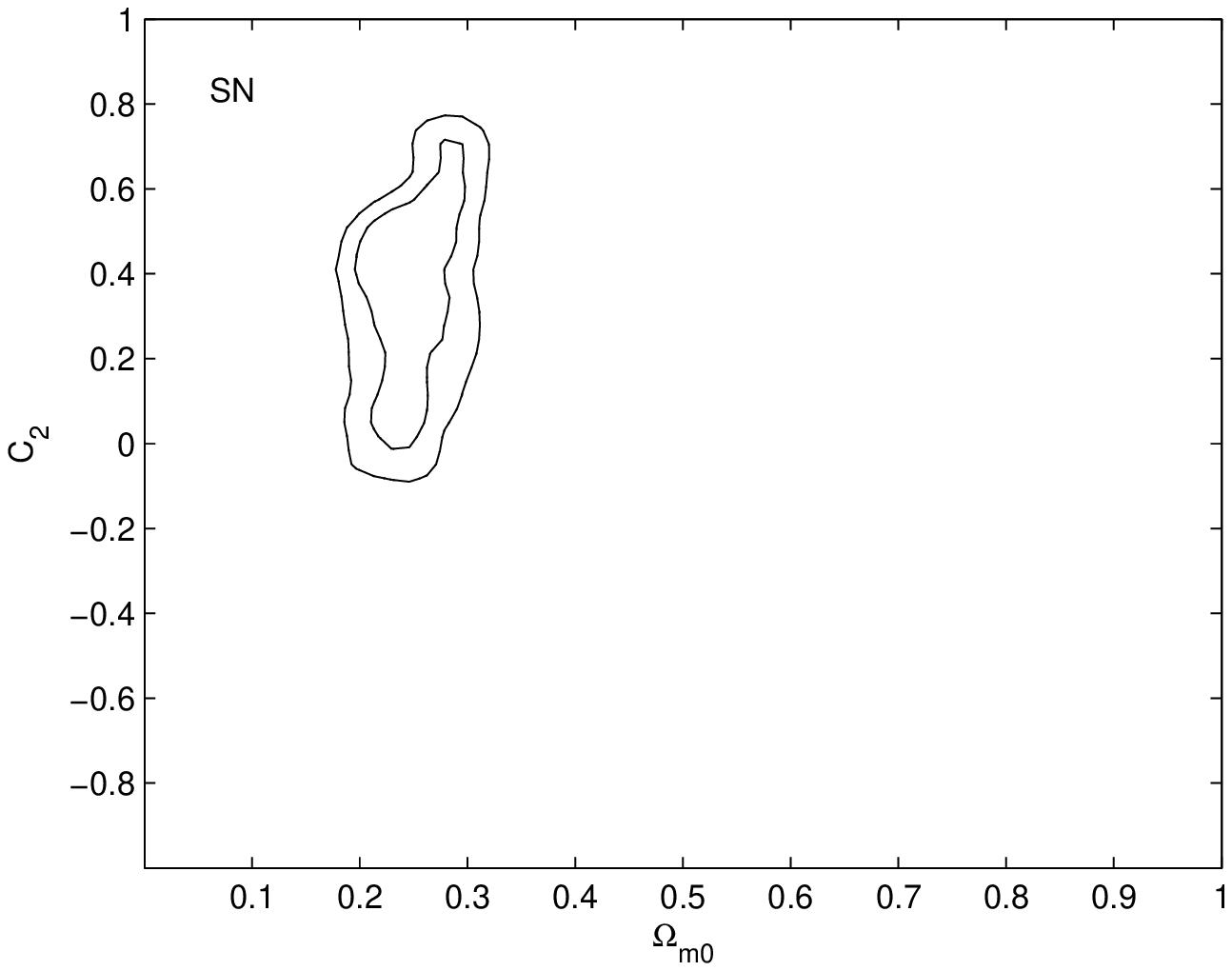}
\includegraphics[scale=0.5]{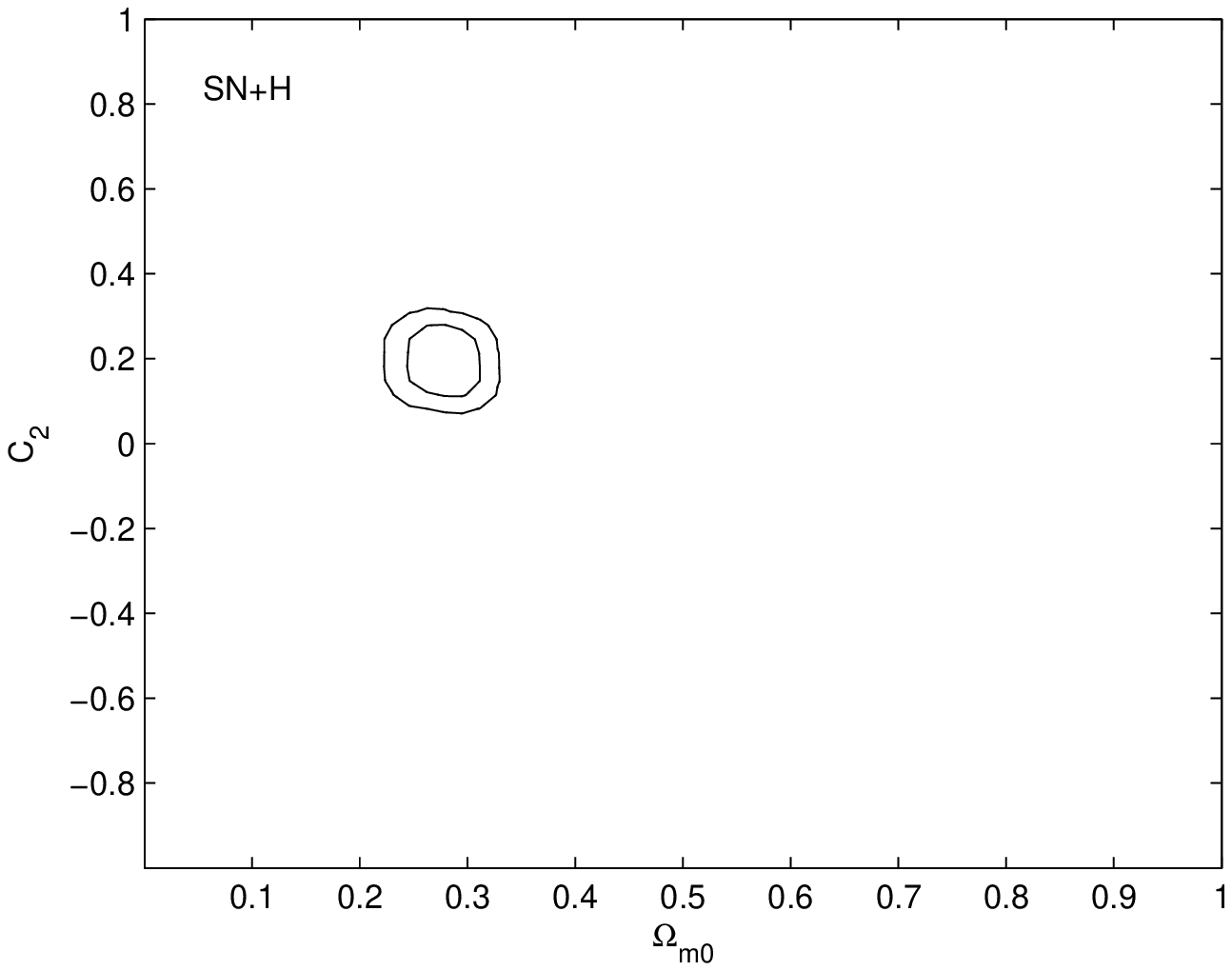}
\includegraphics[scale=0.5]{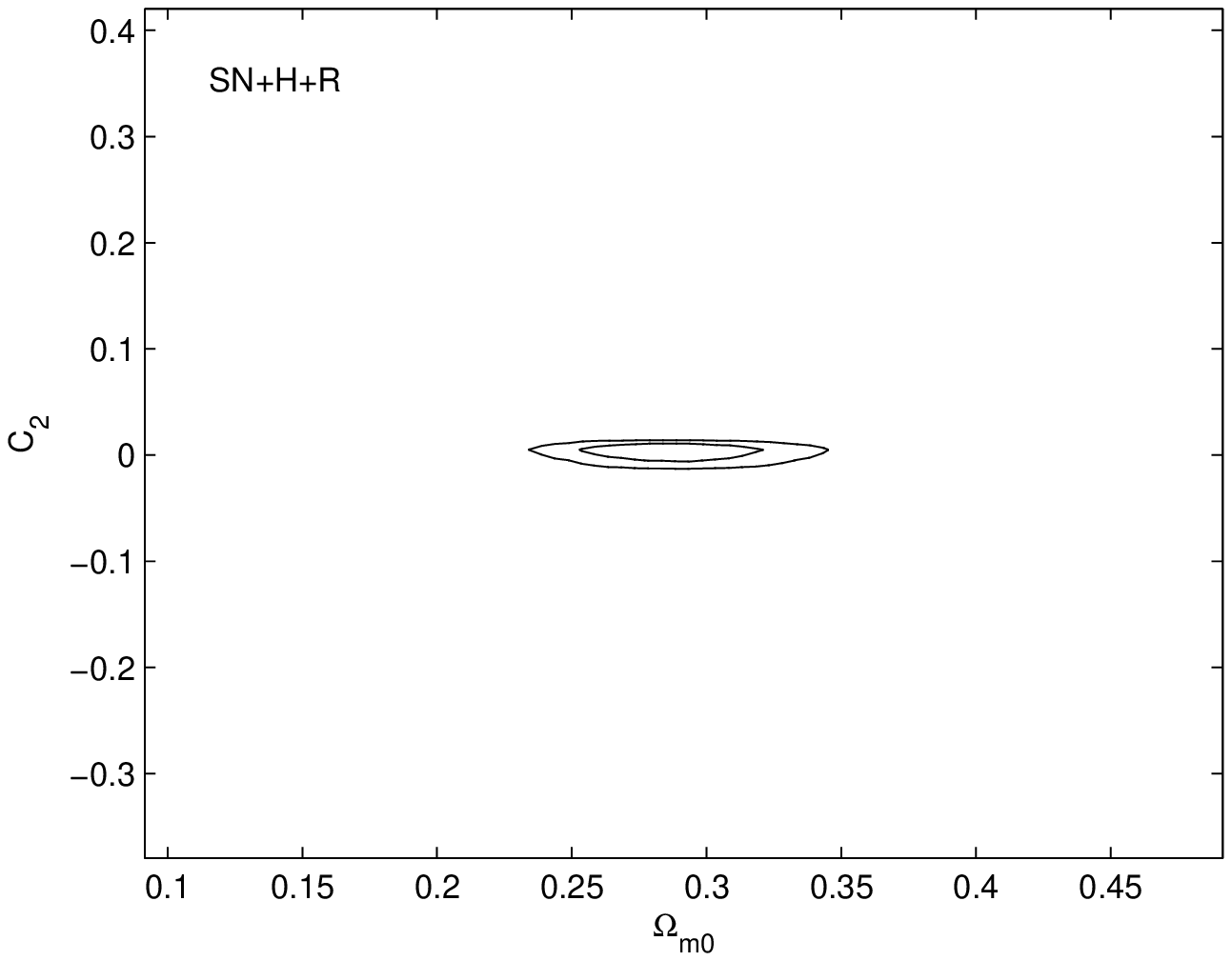}
\includegraphics[scale=0.5]{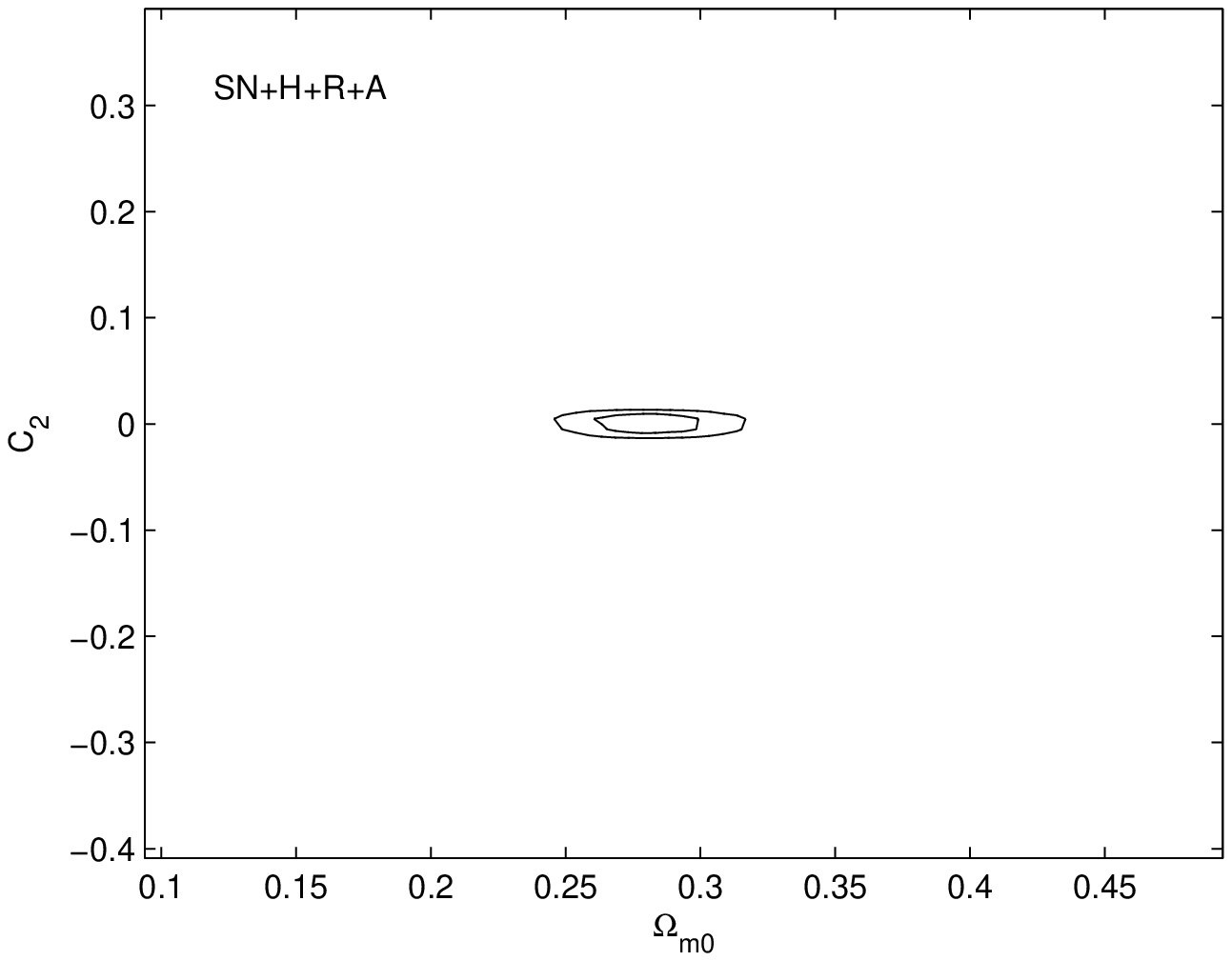}
\caption{Contour plots representing the $68\%$ and $95\%$ credible interval of the joint posterior probability distribution for $\Omega_{m,0}$ and $C_2$ Osc DE 2 model parameters.}
\label{fig:8} 
\end{figure}

To constrain values of parameters for models defined above we repeat the calculation described before. Results are gathered in Table \ref{tab:2} and \ref{tab:3} respectively. Posterior probabilities are presented on Figure \ref{fig:5} and \ref{fig:7} respectively. Two dimensional contour plots in the ($\Omega_{m,0}$, $C_i$) plane are presented on Figure \ref{fig:6} and \ref{fig:8} respectively.

\begin{table}[h]
\centering
\begin{tabular}{c||cccc||cccc|}
\hline
& SN & & & & SN+H & & &  \\
\hline
&Best fit&Mean&$68\%$&$95\%$&Best fit&Mean&$68\%$&$95\%$\\
\hline
$\Omega_{m,0}$&$0.36$&$0.37$&$<0.30,0.44>$&$<0.20,0.48>$&$0.40$&$0.40$&$<0.36,0.45>$&$<0.19,0.48>$\\
$C_1$&$-0.19$&$-0.33$&$<-0.55,-0.07>$&$<-0.92,0.07>$&$-0.34$&$-0.42$&$<-0.70,-0.18>$&$<-0.95,0.11>$\\
\hline
$\chi^2$&$194.99$&&&&$206.75$&&&\\
&&&&&&&&\\
\hline
\hline
& SN+H+R & & & & SN+H+R+A & & &\\
\hline
&Best fit&Mean&$68\%$&$95\%$&Best fit&Mean&$68\%$&$95\%$\\
\hline
$\Omega_{m,0}$&$0.31$&$0.32$&$<0.29,0.34>$&$<0.27,0.38>$&$0.29$&$0.29$&$<0.27,0.31>$&$<0.26,0.32>$\\
$C_1$&$0.97*10^{-6}$&$0.14*10^{-4}$&$<0.66*10^{-6},$&$<0.17*10^{-6},$&$0.42*10^{-6}$&$0.79*10^{-6}$&$<0.40*10^{-6},$&$<0.31*10^{-7},$\\
     &              &              &$0.22*10^{-4}>$ &$0.71*10^{-4}>$ &              &            &$0.12*10^{-5}>$  & $ 0.14*10^{-5}>$ \\
\hline
$\chi^2$&$211.03$&&&&$211.94$&&&\\
&&&&&&&&\\
\hline
\end{tabular}
\caption{Values for oscillating DE 1 model parameters obtained via $\chi^2$ minimization (best fit), values of the mean with the $68\%$ and $95\%$ credible intervals obtained from the posterior probability distribution for considered Osc DE 1 model parameter. }
\label{tab:2}
\end{table}

\begin{table}[h]
\centering
\begin{tabular}{c||cccc||cccc|}
\hline
& SN & & & & SN+H & & &  \\
\hline
&Best fit&Mean&$68\%$&$95\%$&Best fit&Mean&$68\%$&$95\%$\\
\hline
$\Omega_{m,0}$&$0.20$&$0.25$&$<0.22,0.28>$&$<0.20,0.29>$&$0.25$&$0.28$&$<0.26,0.30>$&$<0.24,0.31>$\\
$C_2$&$0.24$&$0.35$&$<0.11,0.54>$&$<-0.03,0.71>$&$0.13$&$0.19$&$<0.14,0.25>$&$<0.12,0.27>$\\
\hline
$\chi^2$&$194.37$&&&&$207.65$&&&\\
&&&&&&&&\\
\hline
\hline
& SN+H+R & & & & SN+H+R+A & & &\\
\hline
&Best fit&Mean&$68\%$&$95\%$&Best fit&Mean&$68\%$&$95\%$\\
\hline
$\Omega_{m,0}$&$0.31$&$0.29$&$<0.26,0.31>$&$<0.24,0.34>$&$0.29$&$0.28$&$<0.26,0.30>$&$<0.25,0.32>$\\
$C_2$&$-0.13*10^{-5}$&$0.82*10^{-5}$&$<-0.40*10^{-5},$&$<-0.85*10^{-5},$&$-0.92*10^{-6}$&$0.25*10^{-6}$&$<-0.11*10^{-5},$&$<-0.23*10^{-5},$\\
     &               &              &$0.24*10^{-4}>$  &$0.32*10^{-4}>$ & & &$0.17*10^{-5}>$ & $ 0.29*10^{-5}>$\\     
\hline
$\chi^2$&$211.03$&&&&$211.94$&&&\\
&&&&&&&&\\
\hline
\end{tabular}
\caption{Values for oscillating DE 2 model parameters obtained via $\chi^2$ minimization (best fit), values of the mean with the $68\%$ and $95\%$ credible intervals obtained from the posterior probability distribution for considered Osc DE 2 model parameter. }
\label{tab:3}
\end{table}

\begin{figure}[h]
\includegraphics[scale=0.85]{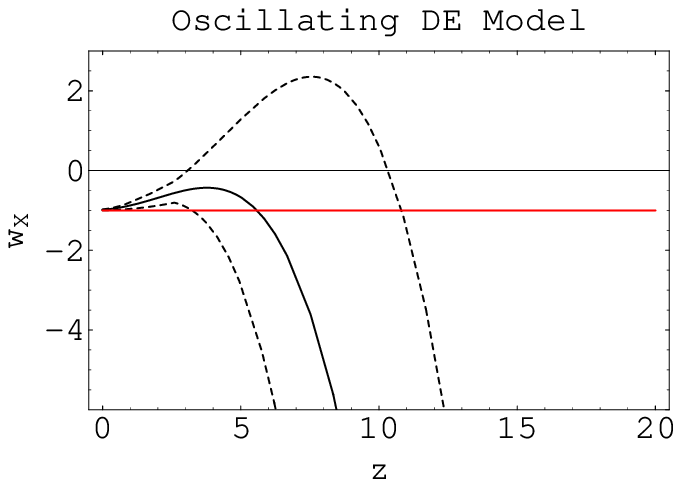}
\includegraphics[scale=0.85]{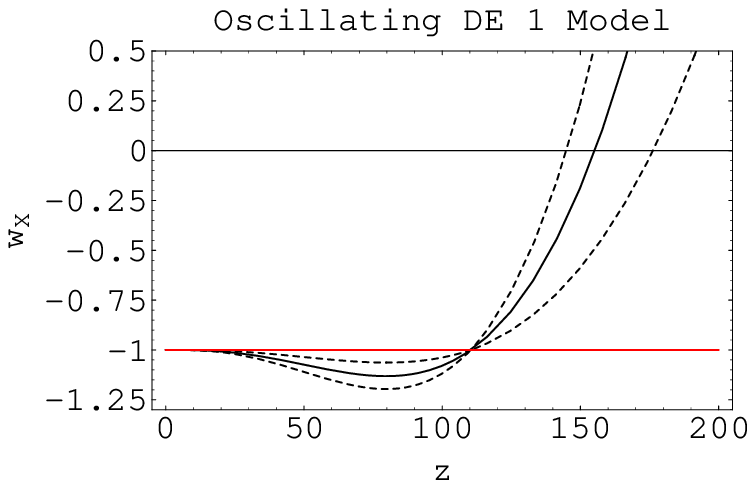}
\includegraphics[scale=0.85]{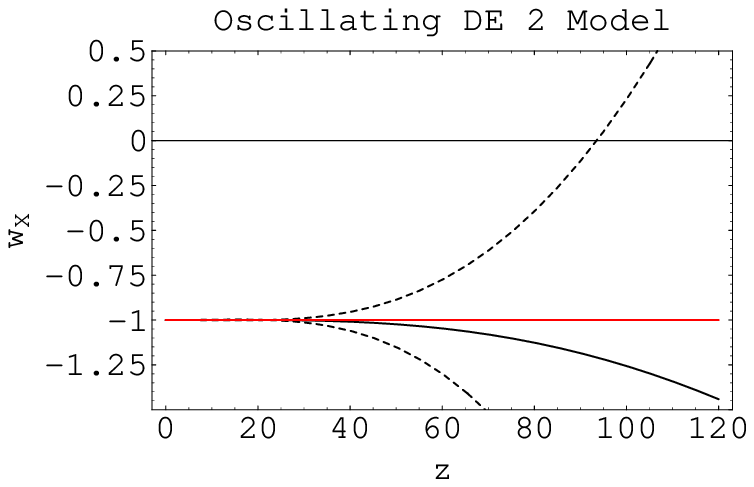}
\caption{$w_X(z)$ for the Osc DE, Osc DE 1, Osc DE 2 Model (black line) together
with $68\%$ credible intervals (dashed lines) (calculated for the
mean and $68\%$ credible interval of the posterior distributions for the model
parameters) and for the  $\Lambda$CDM model (red line).}
\label{fig:9}
\end{figure}

\begin{figure}[h]
\includegraphics[scale=0.85]{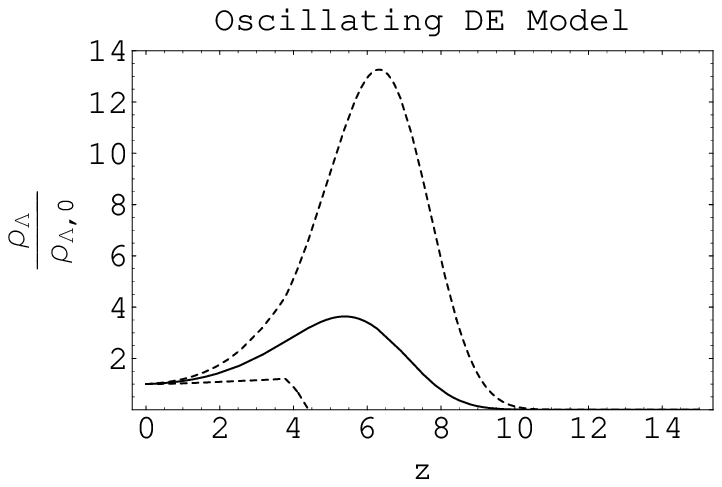}
\includegraphics[scale=0.85]{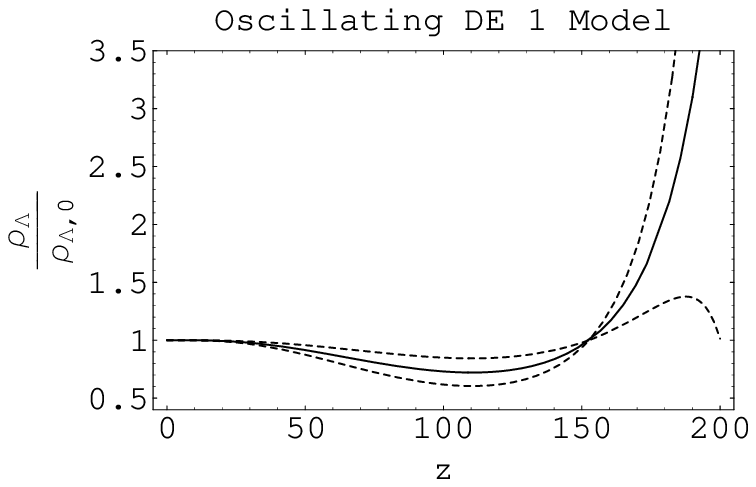}
\includegraphics[scale=0.85]{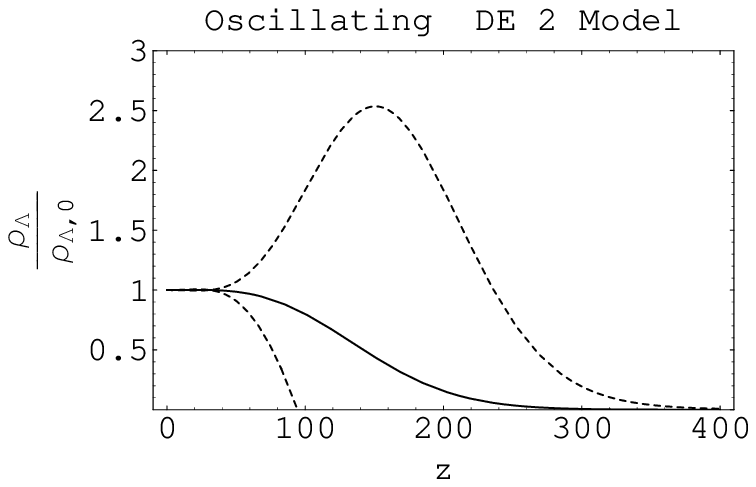}
\caption{$\frac{\rho_{\Lambda}}{\rho_{\Lambda,0}}$ for the Osc DE, Osc DE 1, Osc
DE 2 Model (black line) together with $68\%$ credible intervals (dashed lines)
(calculated for the mean and $68\%$ credible interval of the
posterior distributions for the model parameters).}
\label{fig:10}
\end{figure}

\begin{figure}[h]
\includegraphics[scale=0.85]{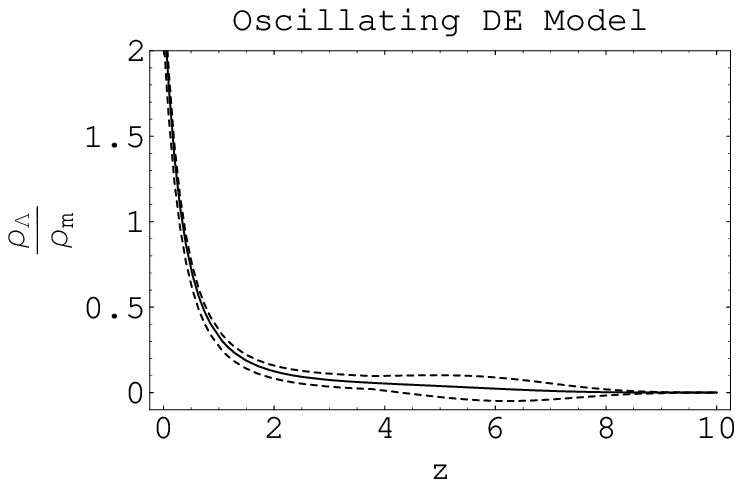}
\includegraphics[scale=0.85]{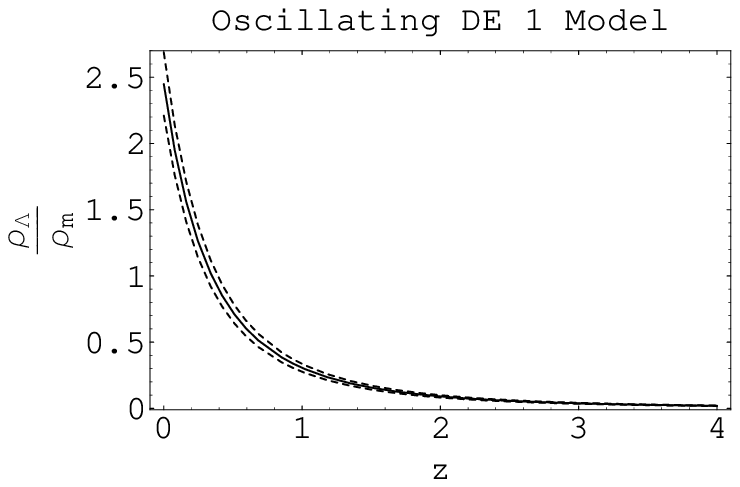}
\includegraphics[scale=0.85]{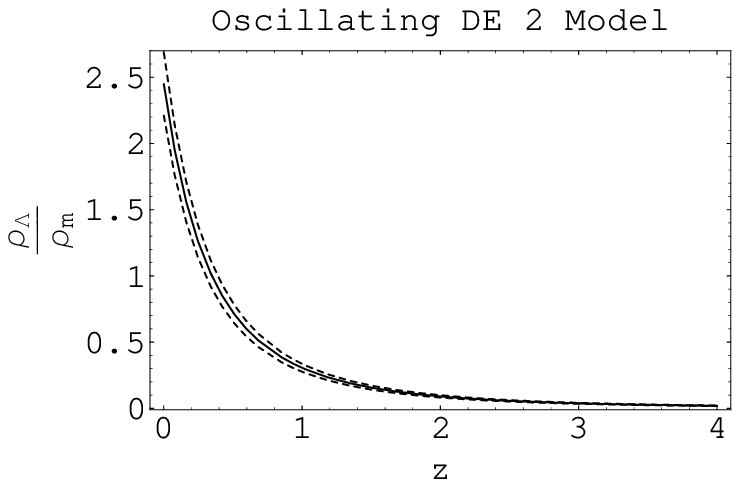}
\caption{$\frac{\rho_{\Lambda}}{\rho_{m}}$ for the Osc DE, Osc DE 1, Osc DE 2
Model (black line) together with $68\%$ credible intervals (dashed lines)
(calculated for the mean and $68\%$ credible interval of the
posterior distributions for the model parameters).}
\label{fig:11}
\end{figure}

The $w_X(z)$, $\frac{\rho_{\Lambda}}{\rho_{\Lambda,0}}$ and
$\frac{\rho_{\Lambda}}{\rho_{m}}$ functions together with $68\%$ credible
interval for considered models ( calculated for the mean of the posterior
distributions for the model parameters which are gathered in Table \ref{tab:1},
\ref{tab:2}, \ref{tab:3} in the SN+H+R+A case) are presented on Figure
\ref{fig:9}, \ref{fig:10}, \ref{fig:11} respectively. \\

Finally we made a comparison of Oscillating DE Models, $\Lambda$CDM model and model with linear in scale factor parametrisation of $w$: $w(a)=w_0+w_1(1-a)$. Analysis was made in the Bayesian framework. Here the best model is this one which has the largest value of the posterior probability. It is convenient to use the posterior odds in analysis, which in the case when no model is favoured a priori is reduced to so called Bayes Factor $B_{ij}$ (the ratio of the evidence for models indexed by $i$ and $j$) \cite{Kass:1995,Szydlowski:2006xx}. This quantity can be interpreted as a strength of evidence against worse model with respect to the better one: $0\leq | 2\ln B_{ij} | < 
2$--not worth more than a bare mention, $2\leq | 2\ln B_{ij} |< 6$ -- positive, 
$6\leq | 2\ln B_{ij} |< 10$ -- strong, and $|2\ln B_{ij}|\ge 10$ -- very strong. Here we used $BIC$ quantity \cite{Schwarz:1978} as an approximation to the minus twice logarithm of the evidence, which is defined in the following way:
\begin{displaymath}
BIC=-2\ln \mathcal{L} +d\ln N, 
\end{displaymath} 
where $\mathcal{L}$ is the maximum of the likelihood function, $d$ is the number
of model parameter and $N$ is the number of data. Values of Bayes Factor
(calculated with respect to $\Lambda$CDM model) are gathered in Table \ref{tab:4}.

\begin{table}[h]
\centering
\begin{tabular}{|c|c|}
\hline
 Model       & $2\ln B$  \\
\hline
             &           \\
$\Lambda$CDM & $0$       \\
             &           \\
Osc DE       & $8.75$    \\
             &           \\
Osc DE 1     & $3.37$    \\
             &           \\
Osc DE 2     & $3.37$       \\
             &           \\
Linear parametrisation & $5.76$ \\
             &        \\
\hline          
\end{tabular}
\caption{Twice logarithm of the Bayes Factor}
\label{tab:4}
\end{table}

As one can conclude $\Lambda$CDM model is the best one from the set of models considered in this paper. Evidence in favour this model is strong when comparing with the Osc DE model and positive in the other cases. There is positive evidence in favour model with Linear parametrisation over the Osc DE model. Bayes Factor computed for Osc DE 1 and Osc DE 2 models is close to $1$ which indicate that the information coming from the data (used in analysis) are not enough to favour one of this model over another. In this situation calculation the Bayesian evidence by numerical integration could give better results. Finally Osc DE 1 and Osc DE 2 are favoured over the Osc DE Model and over model with linear in $a$ parametrisation of $w$.

\section{Conclusions}

In this paper we have placed constraints on a parametrised dark energy model
\cite{Hrycyna:2007mq} using the SNIa data sets, observational H(z) data, the size of the
baryonic acoustic oscillation peak from SDSS and the shift parameter from the
CMB observations. We study possibility that phantom dark energy is oscillating
rather than decaying to $\Lambda$. Such a scenario opens the possibility of the
non-minimal coupling to gravity for phantom scalar field. Combining four data
bases (SNIa, H(z), CMB, SDSS) we obtain constraints on the oscillating dark energy model parameters
$(\Omega_{m},D_1,D_2)$ and compare this model with $\Lambda$CDM model and with model with linear in $a$ parametrisation of $w$ in the Bayesian framework. It is found that special cases of oscillating phantom dark energy model ( called Osc DE 1 and Osc DE 2 in this paper) are favoured
over the model with linear in $a$ parametrisation of $w$. Cosmological constant case still remains as the best one from the set of considered models.\\

\begin{acknowledgments}
This work has been supported by the Marie Curie Actions
Transfer of Knowledge project COCOS (contract MTKD-CT-2004-517186).
\end{acknowledgments}

\end{document}